\documentclass[10pt]{cup-elements}

\usepackage{epsfig}
\usepackage{amsmath,amssymb,bm}
\usepackage{graphics}
\usepackage{epsfig}
\usepackage{graphicx}
\usepackage{verbatim}
\usepackage{xcolor}

\def\ER{Erd\H{o}s-R\'enyi }
\usepackage[natbibapa]{apacite}
\CUPseries{Series Name}
\CUPelements{Elements Name}

\title{Weak Multiplex Percolation}

\author{G.~J.~Baxter}

\affil{Departamento de F{\'\i}sica da Universidade de Aveiro $\&$ I3N, Campus Universit\'ario de Santiago, 3810-193 Aveiro, Portugal}

\author{R.~A.~da~Costa}
\affil{Departamento de F{\'\i}sica da Universidade de Aveiro $\&$ I3N, Campus Universit\'ario de Santiago, 3810-193 Aveiro, Portugal}

\author{S.~N. Dorogovtsev}
\affil{Departamento de F{\'\i}sica da Universidade de Aveiro $\&$ I3N, Campus Universit\'ario de Santiago, 3810-193 Aveiro, Portugal}

\author{J.~F.~F. Mendes}
\affil{Departamento de F{\'\i}sica da Universidade de Aveiro $\&$ I3N, Campus Universit\'ario de Santiago, 3810-193 Aveiro, Portugal}

\begin{document}

\frontmatter  
\maketitle

\begin{abstract}
In many systems consisting of interacting subsystems, the complex interactions between elements can be represented using multilayer networks. However percolation, key to understanding connectivity and robustness, is not trivially generalised to multiple layers.
We describe a generalisation of percolation to multilayer networks: weak multiplex percolation. 
A node belongs to a connected component if at least one of its neighbours in each layer is in this component. 
We fully describe the critical phenomena of this process. In particular, in two layers, with finite second moments of the degree distributions, an unusual continuous transition with quadratic growth above the threshold. When the second moments diverge, the singularity is determined by the asymptotics of the degree distributions, creating a rich set of critical behaviours. 
In three or more layers we find a discontinuous hybrid transition which persists even in highly heterogeneous degree distributions, becoming continuous only when the powerlaw exponent reaches $1+ 1/(M-1)$ for $M$ layers.
\end{abstract}



\copyrightauthor{Baxter, da Costa, Dorogovtsev, Mendes, 2020}

\mainmatter  

%


\section{Introduction}

Many complex systems consist of multiple interacting
sub-systems. 
Examples include financial \citep{caccioli2014stability, huang2013cascading}, infrastructure
\citep{rinaldi2001identifying}, informatic \citep{leicht2009percolation} and ecological \citep{pocock2012robustness} systems.
To understand the
functioning of one sub-system or layer, it is necessary to take into
account dependencies upon and interactions with the other sub-systems,
which can significantly alter the behaviour of the system \citep{buldyrev2010catastrophic}.

A convenient and powerful representation for many such
systems is as a set of interdependent networks, with each subsystem
represented by a separate network layer. Interdependencies are represented as special dependency links between nodes in different layers. The presence of a dependency link indicates that the failure of a node in one layer will lead to the failure of nodes in other layers to which it is connected by such dependency links.
These interdependencies may dramatically increase the fragility of the system
\citep{buldyrev2010catastrophic,baxter2012avalanche}. The dependency may be full, so that every node in one layer is interdependent with a partner node in each other layer, or partial, so that some nodes have dependencies in only some of the other layers \citep{dong2012percolation}.

In many cases, a common set
of nodes can be defined across all layers, allowing a multiplex
network representation \citep{son2012percolation}, which simplifies the analysis. That is, if the dependency connections are one-to-one, we may merge interdependent nodes, as they will always be removed together. The different layers then consist of different types (colours) of connections between the same set of nodes. Such a network, one set of nodes with different types of edges between them, is called a multiplex network. Note that this mapping is still possible even when the multilayer network is only partially interdependent, or when not all nodes appear in every layer. Nodes which do not appear in a given layer have no connections of the corresponding colour in the multiplex representation. 

In a single network, a giant connected component, containing a finite fraction of all nodes in the network, appears at a well defined percolation threshold with respect to a control parameter affecting the density of the network (mean degree, or fraction of nodes or edges surviving random damage, for example). This transition is typically a second-order continuous transition, with the relative size of the giant component growing linearly with the distance above the threshold, although this growth exponent may be strongly affected if the degree distributions is very broad \citep{cohen2002percolation, dorogovtsev2008critical}.

\begin{figure}[t]
\centering
\includegraphics[width=0.9\textwidth]{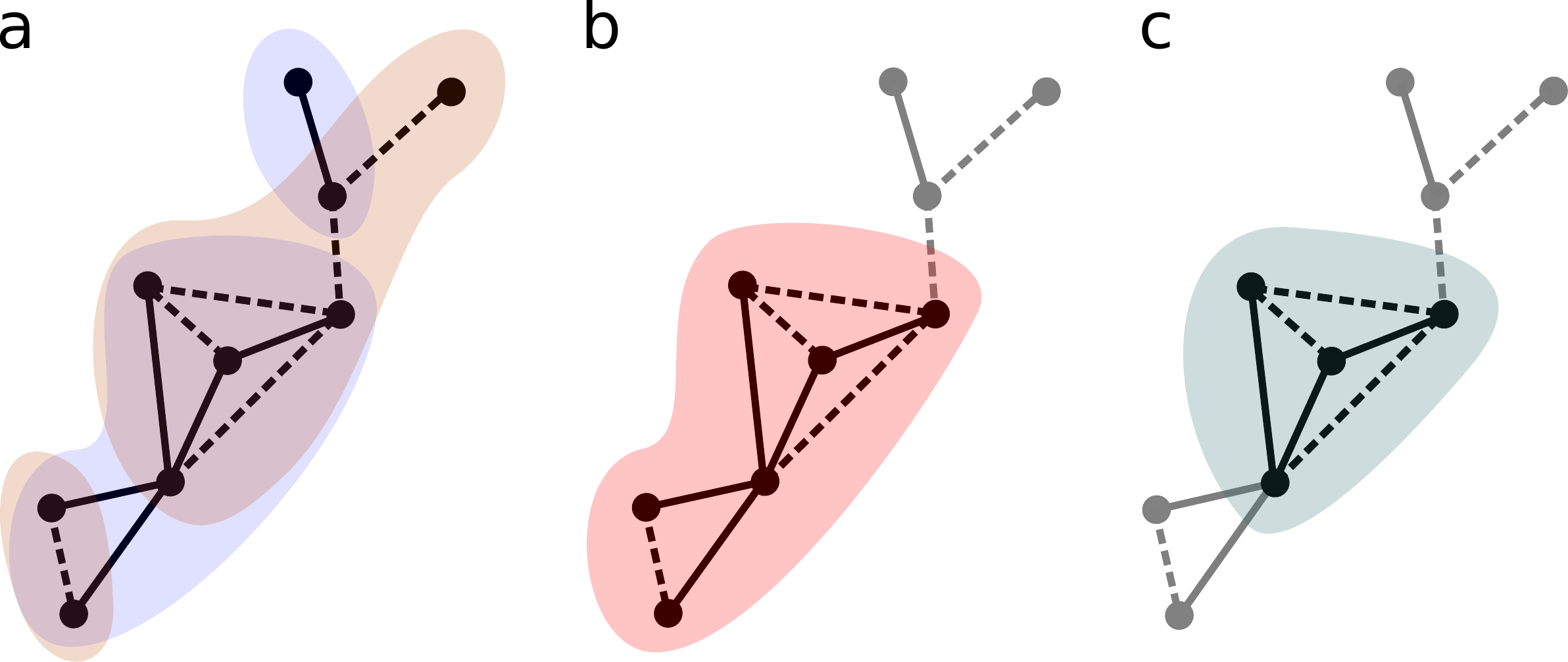}
\caption{Comparison between percolation generalisations in multiplex networks. a) A two layer multiplex network, with connected clusters in each layer shaded. b) A connected cluster under weak multiplex percolation. Nodes must have at least one connection of each type to at least one other member of the cluster. Nodes do not necessarily belong to the same cluster in each layer when considered separately. c) A mutually connected cluster. Nodes must have a path to all other members of the cluster in both layers, i.e. they must belong to the same connected cluster in both layers.}
\label{fig_percolation_rules}       
\end{figure}

The percolation problem may be generalised to multiplex or multilayer networks in two ways:

(i) In the percolation process described in \citet{buldyrev2010catastrophic,son2012percolation,baxter2012avalanche}, a percolating cluster consists of a set of nodes, each pair of which is connected  by a path in every layer of the multiplex network to which they both belong, see Figure \ref{fig_percolation_rules} (c). Such clusters are referred to as mutually connected clusters. 
One finds a greatly increased fragility of the system. Failures of nodes in one layer lead to  failures in another layer, affecting the connectivity of other nodes in that layer, which may then fail, causing further failures in the first layer. In this way, damage may cascade back and forth between layers. A small initial damage may lead to a discontinuous collapse of the giant mutually connected cluster. The collapse is a discontinuous hybrid phase transition, which differs from a first-order transition in that one finds a square root singularity above the transition, with diverging susceptibility. The phase transition therefore has some properties in common with second-order transitions. The same type of transition is observed in $k$-core percolation \citep{dgm2006}.
A consequence of this definition is that in order to establish whether a given node belongs to a percolating cluster, one must explore the whole cluster to which it belongs, in all layers in which it is present. One may identify the giant mutually connected component by iteratively removing all finite connected components in each layer until an equilibrium is reached.
Finding the giant mutually connected cluster is therefore a computationally intensive process, as compared with, say, $k$-core pruning \citep{baxter2015critical}, in which a local pruning rule may be applied to find the giant $k$-core cluster.

Since its proposal by \citet{buldyrev2010catastrophic}, significant attention has been devoted to this process, exploring the effects of partial interdependence \citep{dong2012percolation}, multiple dependencies \citep{shao2011cascade}, correlations \citep{hu2013percolation} and overlapping edges \citep{min2015link,baxter2016correlated,cellai2016message} among many others \citep{bianconi2018multilayer,kivela2014multilayer,boccaletti2014structure,cozzo2018multiplex}.
The mutually connected component in multiplex networks is strongly affected by highly heterogeneous network structure \citep{baxter2012avalanche}. In two layers with powerlaw-tailed degree distributions, one finds that both the height of the discontinuity and the critical point tends to zero as the powerlaw exponent tends to $\gamma=2$ from above.

(ii) 
An alternative percolation rule was proposed by \citet{baxter2014weak}. Under this rule, 
nodes are considered active if they maintain at least one connection to another active node in each layer to which it belongs. Connected clusters are formed from such active nodes, with two nodes belonging to the same cluster if there is a path between them ignoring to which layer the edges belong.
a node belongs to a given cluster if it is connected to at least one member of the cluster in every layer to which it belongs. There is not necessarily a path between every pair of nodes in the cluster in every layer, see Figure \ref{fig_percolation_rules} (b).
Under this rule, the giant component may be identified by iteratively applying a local pruning process, removing any nodes without the required connections, without needing to repeatedly identify the full clusters. The computation required is similar to that of $k$-core pruning. 
This problem is referred to as weak multiplex percolation, to distinguish it from the more restrictive rule of the mutually connected cluster.

This process was further explored in \citep{min2014multiple}, and the relationship with the stronger rule elaborated in \citep{baxter2016unified}. A complete description of the critical phenomena associated with weak multiplex percolation was given in \citep{baxter2020weak}. In two-layer networks the problem is equivalent to $(1{-}1)$-core percolation, as proposed in \citep{azimi2014k}. One observes a continuous transition, but typically with quadratic (rather than linear) growth above the critical point. In three or more layers hybrid phase transition occurs, of the same type as for the mutually connected components (rule (i) above). 


The choice of which multiplex percolation definition is appropriate depends on the problem in hand.
In many situations, the functioning or survival of an agent or site is only dependent on its relations with close neighbours---support and distribution of distinct resources or information, for example \citep{min2014multiple}. For such problems, weak multiplex percolation applies. 
For systems requiring connection to a common or centralised system, such as electricity supply and control, the mutually connected clusters might be more appropriate \citep{buldyrev2010catastrophic}.
Connected components for these two percolation formulations are compared in Figure \ref{fig_percolation_rules}.

The unusual phase transitions observed in both these problems, a hybrid transition which is discontinuous like a first order transition, and a square root singularity on one side of the transition, with diverging susceptibility like a second order transition, 
have also been observed in $k$-core percolation \citep{dgm2006}, in the activation process of bootstrap percolation \citep{baxter2010bootstrap} and the Kuramoto synchronisation model \citep{moreno2004synchronization}, as well as in activation
processes on multiplex and multilayer networks
\citep{baxter2014weak,min2014multiple} among others.
What these processes have in common is the possibility for avalanches. A change in the status of one node may alter the status of a neighbouring node, which in turn may affect further nodes. For example, in the $k$-core problem, nodes belong to the $k$-core if they have at least $k$ neighbours within the core. If a node is removed, its neighbours lose a connection, and may then have less than $k$ connections, so are themselves removed. These chains or avalanches of removals can be seen as a branching process. The phase transition occurs when the branching ratio exceeds one, so each removal, on average, leads to more than one further removal. The avalanches diverge in size and consume a finite fraction of the system, leading to a discontinuity in the system size \citep{baxter2015critical}.

Our aim in this Element is to summarise the behaviour observed under weak multiplex percolation, the methods used to calculate the size of the giant component, and highlight the unusual critical phenomena observed in this problem.
We detail the effects of heterogeneous degree distributions, showing that, in contrast with other network percolation problems, the discontinuity and critical point remain nonzero at $\gamma=2$, finally vanishing at $\gamma = 1+1/(M-1)$, where $M$ is the number of layers. These phenomena are placed in context with related percolation problems.
The results presented here are based largely on those presented in \citep{baxter2014weak} and \citep{baxter2020weak}.

The remainder of this Element is organised as follows. In the next Section we precisely define the problem, and give the self consistency equations which allow its solution. We compare weak multiplex percolation with ordinary percolation and with the mutually connected component in Section \ref{compare}.
 In Section \ref{rapidlyDecaying} we derive the main results for rapidly decaying degree distributions. The effect of heterogeneous degree distributions is then detailed in Section \ref{heavyTailed}, including the disappearance of the hybrid transition for very slowly decaying degree distributions. Final discussion is given in Section \ref{conclusions}.


\section{Weak multiplex percolation}\label{weakpercolation}

For the purposes of this Element, we will assume a multiplex formulation of the problem.
A multiplex network consists of a set of $N$ nodes, connected in $M>1$ layers, each with its own type (colour) of edge. A node may be connected in all $M$ layers, or only a subset of them.
In a simple network (i.e. $M=1$ layer), two nodes belong to the same connected component if there is a path between them following the edges of the network. 
One may vary the density of connections in the network, for example by changing the mean degree, or occupying edges or nodes with probability $p$. 
When the network is very sparse, with few connections, connected components are small. As we increase the density of connections, connected clusters grow and may start to join together to form larger clusters. Eventually the largest connected component may contain a significant fraction of all the nodes in the network.
In the infinite size limit $N\to \infty$, connected components are either finite (containing a finite number of nodes) and thus occupying a vanishing fraction of the whole network, or giant, occupying a nonzero fraction of the network.
Beginning with a low density, only finite clusters exist. As we increase the network density, 
a giant connected component appears at a well defined threshold, above which the giant component grows linearly for sufficiently rapidly decaying degree distributions. This is the classic percolation transition.

In weak multiplex percolation, a node is active if it maintains a connection in each layer to other active nodes. Two active nodes are part of the same weak percolation component if there is a path between them via edges in any layer (i.e. ignoring edge colours).
To identify the weak percolation clusters in a given multiplex network, one may simply prune any nodes that do not have connections in all layers (inactive nodes) in which they participate, repeating the pruning until an equilibrium is reached. The connected clusters in the projection of the remaining network are then the weak percolating clusters \citep{baxter2014weak,baxter2016unified}.

\begin{boxedtext}{Identifying weak multiplex percolation clusters}
In a given multiplex network, one may use the following algorithm to identify the weak percolating clusters:
\begin{enumerate}
\item Set the status of all nodes in the network to active
\item For each node $i$, for each layer $s$ in which $i$ has degree $>0$, check whether $i$ has at least one active neighbour in $s$. If not, set the status of $i$ to inactive.
\item Repeat step 2. until no changes to node statuses are made.
\item Identify connected clusters within the subnetwork consisting only of active nodes and the edges between them. These are the weak percolating clusters.
\end{enumerate}
\end{boxedtext}

 In a two layer network with sufficiently rapidly decaying degree distribution (such as Poisson degree distributions, as found in  \ER networks) the giant component appears continuously with a second-order phase transition and grows as the square of the distance from the critical point. This differs from the usual percolation transition, which exhibits linear growth. The mutually connected component never appears with a continuous transition in such networks, for any number of layers. 
For three or more layers, the giant component appears with a discontinuous hybrid transition, of the same kind found in the mutual connected component of multiplex networks \citep{baxter2012avalanche}, and in $k$-core percolation \citep{dgm2006,baxter2011heterogeneous}. The size of the giant component $S$ jumps from zero to a finite value at the critical point, and there is a square-root singularity above the transition.

\begin{figure}
\centering
\includegraphics[width=0.8\textwidth]{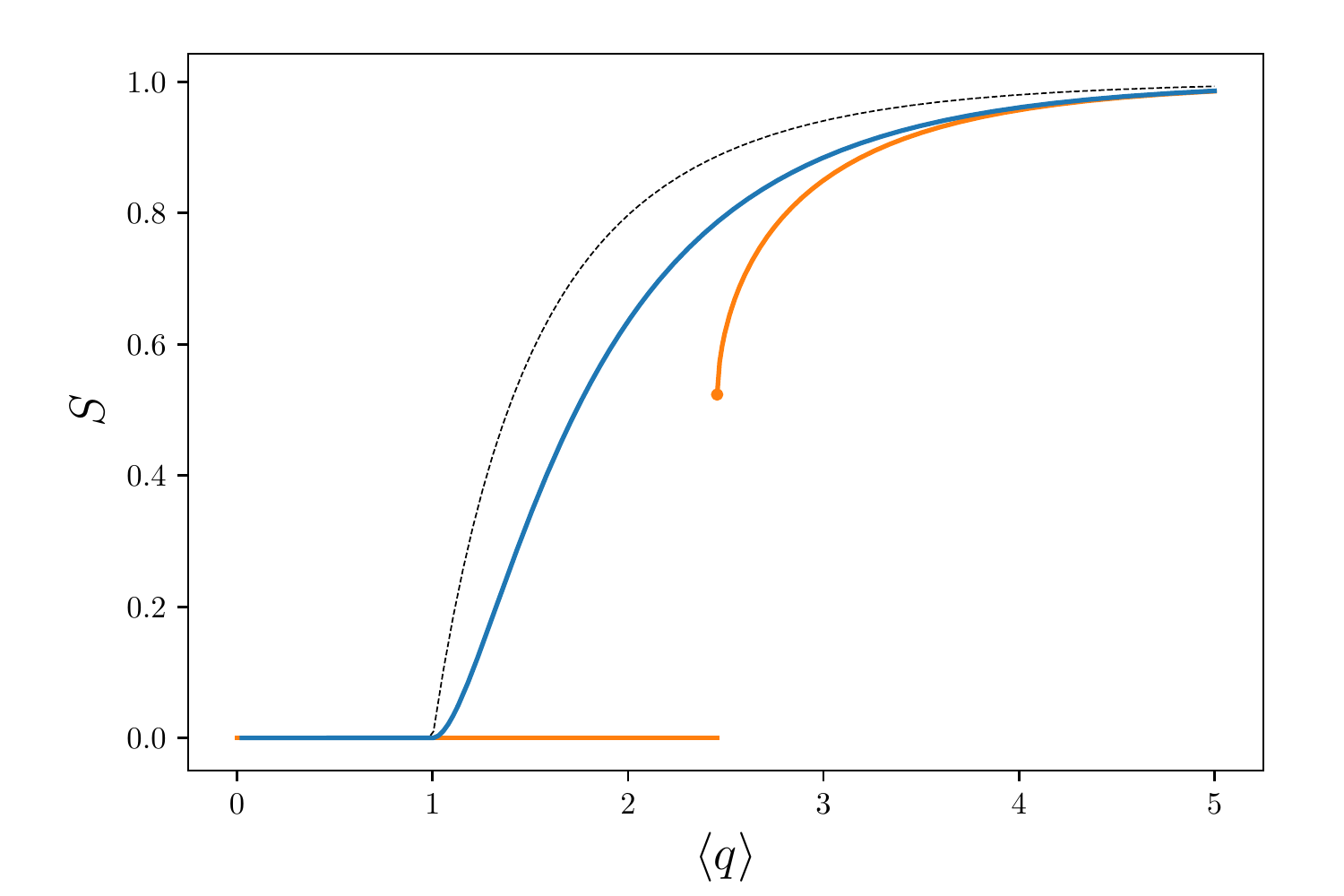}
\caption{Relative size of the weak multiplex percolation giant component as a function of mean degree in \ER networks, for $M=2$ layers (blue), showing the continuous phase transition, and $M=3$ layers (orange) showing the discontinuous hybrid transition. For comparison, the relative size of the single layer \ER network giant component is also shown (dashed line).}\label{rapidlyExamples}
\end{figure}


\subsection{Self consistency equations for weak multiplex percolation}\label{selfconsistencyWP}

 To understand the critical behaviour of weak multiplex percolation, let us consider a large sparse random multiplex network, consisting of $N$ nodes connected in $M$ layers (colours), each having its own unique type of edge. This is equivalent to considering a network with $M$ different types of edges connecting the nodes. For clarity, we refer to the different types of edges as different {\em colours}. Note that a node does not necessarily participate in all layers. Nodes which initially have no connections in a given layer are considered not to participate in that layer.
 A node is considered active if it maintains at least one connection to another active node in each of the layers in which it participates. A weak percolating cluster is then a set of such active nodes which are connected to each other (each member is connected to at least one other member in at least one layer).

 We consider a generalised configuration model, defined by its joint degree distribution $P(q_1,q_2,...,q_M)$. This allows for arbitrary degree correlations between layers, which do not at all impede the analysis.
Each layer $l$ is therefore a random graph, defined by it's internal degree distribution $P(q_l) \equiv \sum_{i\neq l}\sum_{q_i}P(q_1,q_2,...,q_M)$ . This formulation does not however consider degree-degree correlations within layers, although  one may generalise the analysis to consider them.

As the number of nodes $N$ tends to infinity, the relative prevalence of finite loops in each layer tends to zero, and each layer can be considered locally treelike. This property allows us to write self consistency equations to calculate the relative size of the giant weak percolation cluster. 
The advantage of dealing with a treelike network is that we may consider the connectivity of each neighbour of a given node to be independent of the other neighbours. Furthermore, in a configuration model network with no neighbour degree correlations, an edge emanating from a randomly selected node of degree $q$ leads to node of degree $q'$ with probability $q'P(q')/\langle q\rangle$.

\begin{figure}
\centering
\includegraphics[width=0.8\textwidth]{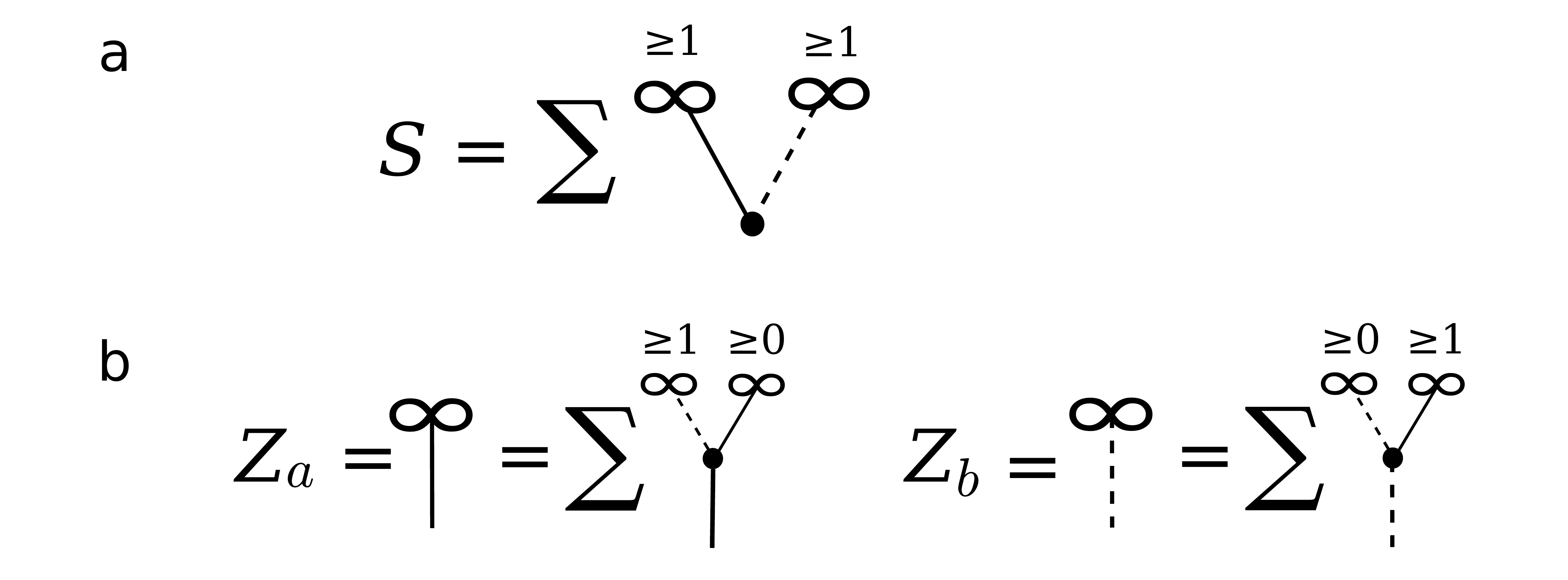}
\caption{
Diagrammatic representation of self-consistency
equations in a multiplex network with $M = 2$ tree-like layers, labelled $a$ and $b$. (a) A node belongs to the giant weak percolation cluster
(giant component) if it has at least one connection via an
edge of type $a$ satisfying the configuration which occurs with probability
$Z_a$ (represented by a solid edge leading to an infinity symbol), and one of type $b$ satisfying $Z_b$ (dashed
edge leading to an infinity symbol). This corresponds to Equation (\ref{S_general}). 
(b) The probabilities $Z_a$ (left) and $Z_b$ (right) obey recursive relations corresponding to Equation (\ref{Z_general}). For $Z_a$, an edge in layer $a$ leads to
a node with at least one outgoing edge of type $b$ satisfying $Z_b$, and similarly for $Z_b$.
}\label{selfConsistency}
\end{figure}

We can then write a self consistency equation for the probability $S$ that a randomly selected node is active, that is, that it belongs to a weak percolation cluster. This occurs if in each layer $l=1,...,M$ the node has at least one neighbour which fulfils the required configuration, which we will define in a moment, and which occurs with probability $Z_l$ in layer $l$. This condition is represented diagrammatically for two layers in Figure \ref{selfConsistency} a.
Since the probability for each neighbour is independent, the probability that it occurs in at least one neighbour is simply a sum of binomial factors. We can thus write:
\begin{align}
S &= 
\sum_{q_1,q_2,...,q_M}\!\!\!\!\!\!
P(q_1,q_2,...,q_M)
\prod_{l = 1}^M
\sum_{m=1}^{q_l} \binom{q_l}{m} Z_l^{m} (1-Z_l)^{q_l-m}
\nonumber\\
&= \sum_{q_1,q_2,...,q_M}\!\!\!\!\!\!
P(q_1,q_2,...,q_M)
\prod_{l = 1}^M
[1-(1-Z_l)^{q_l}]. \label{S_general}
\end{align}
To calculate the probabilities $Z_l$, one may use a recursive argument similar to the one above. Let us imagine following a randomly selected edge in layer $l$ to one of its ends. Clearly the node we reach has a connection in layer $l$. It also has connections in each of the other layers with probabilities given by factors of the same form as in the equation for $S$. Thus we can write an equation for $Z_l$ purely in terms of the probabilities $Z_m$:
 \begin{align}\label{Z_general}
Z_l &= 
\sum_{q_1,q_2,...,q_M}\!\!\!\!\!\!
\frac{q_l P(q_1,q_2,...,q_M)}{\langle q_l\rangle}
\prod_{m \neq l}
[1-(1-Z_m)^{q_m}]
\nonumber \\
&\equiv \Psi_l(Z_1,...,Z_{l-1},Z_{l+1},...,Z_M)\,
\end{align}
for $l = 1,2,...,M$.  We define the right hand side of this equation as the function $\Psi_l(Z_1,...,Z_{l-1},Z_{l+1},...,Z_M)$.
These self consistency equations are represented diagrammatically, for two layers, in Figure \ref{selfConsistency} (b).

One may also write this equation using a vector notation
\begin{align}\label{Z_general_vector}
\bf{Z} = \bf{\Psi}(\bf{Z})\,,
\end{align}
where $\bf{Z}$ is a vector whose elements are $Z_1,Z_2,...,Z_M$, and $\bf{\Psi}$ is a vector function, $\Psi_i$, $i \in \{1,...,M\}$, of these variables.

As long as the total degree  $q_1+q_2+...+q_M$ is much smaller than the number of nodes in the network, that is, the projected network is sparse, the relative frequency of finite loops in this projected network is also vanishing. This means that the probability of a node belonging to a finite weak percolation cluster tends to zero. Thus the probability $S$ is also the probability that a node belongs to a giant weak percolation cluster (which we will refer to henceforth simply as the ``giant component"). This is the same as the relative size of the giant component.
One may thus obtain the size of the giant percolating cluster by first solving simultaneously the recursive Equations (\ref{Z_general}), then substituting into Equation (\ref{S_general}).

Consider for a moment the simplest case of a symmetric uncorrelated multiplex network, in which all layers have the same degree distribution, with no degree correlations between layers. Then we need only a single variable $Z$, which is the same for all layers. In this case, in $M=2$ layers $\Psi(Z)$ is a concave function of $Z$, meaning that the weak percolation giant component appears with a continuous transition, while in three or more layers it is a convex function, and the transition becomes discontinuous. We demonstrate these critical phenomena in detail in Section \ref{rapidlyDecaying} below.


	\section{Relation to other percolation models}\label{compare}

At this point it is instructive to compare the weak percolation model with two related models: percolation in a single layer network, and the mutually connected component in multiplex or multi-layer networks. Both may be examined through the use of self consistency equations, in a very similar way to what we have already done for weak multiplex percolation.  A brief summary of these two models highlights the differences in the critical behaviour, and hence the unique properties of weak multiplex percolation.

\subsection{Percolation in a single layer network}
	
In the limit of a single layer, $M=1$, both weak multiplex percolation and the mutually connected component coincide with the usual percolation in a network. Two nodes belong to the same cluster if there is at least one path between them.
One may use the mean degree of a random network as a control parameter, or alternatively one may apply random damage to a given network, retaining a fraction $p$ of nodes (site percolation) or of all edges (bond percolation) in the network.
A giant connected component appears at a critical value of the control parameter, and typically grows linearly with the distance above the critical point, although nonlinear exponents may appear in strongly heterogeneous networks \citep{dorogovtsev2008critical}.

In large sparse uncorrelated random networks, one may use the tree ansatz to write self consistency equations for the relative size of the giant connected component, just as we have done for weak multiplex percolation.
A node belongs to the giant component if it has at least one edge leading to an infinite subtree, which occurs with probability $X$:
\begin{align}\label{S_ordinary}
S = \sum_q P(q) \left[1 - (1-X)^{q} \right]
\end{align}	
where $X$ obeys the recursive equation
\begin{align}\label{X_ordinary}
X = \sum_q \frac{qP(q)}{\langle q\rangle} \left[1 - (1-X)^{q-1} \right]\,,
\end{align}
compare Equation (\ref{Z_general}).

Linearising Equation (\ref{X_ordinary}) we find the criterion for the critical threshold:
\begin{align}\label{thresh_ordinary}
\langle q\rangle = { \langle q(q-1)\rangle}.
\end{align}

Expanding Equation (\ref{X_ordinary}) for small $X$ and now keeping terms up to second order allows us to find the behaviour near the critical point. We find:
\begin{align}\label{X_2nd_ordinary}
 X = \frac{\langle q(q-1)\rangle  - \langle q\rangle }{ \langle q(q-1)(q-2)\rangle }
\end{align}
and thus, using the expansion of Equation (\ref{S_ordinary}),
\begin{align}\label{S_ordinary}
S = \frac{\langle q\rangle[\langle q(q-1)\rangle  - \langle q\rangle] }{ \langle q(q-1)(q-2)\rangle }.
\end{align}	
Considering, for example, the mean degree $\langle q\rangle$ as a control parameter, and comparing with Equation (\ref{thresh_ordinary}) we see that the giant component grows linearly above the critical point, see Figure \ref{rapidlyExamples}.
This is the classical percolation transition: a continuous second-order transition with linear growth above the critical point.


	\subsection{The mutually connected component in multi-layer networks}

\begin{figure}
\centering
\includegraphics[width=0.8\textwidth]{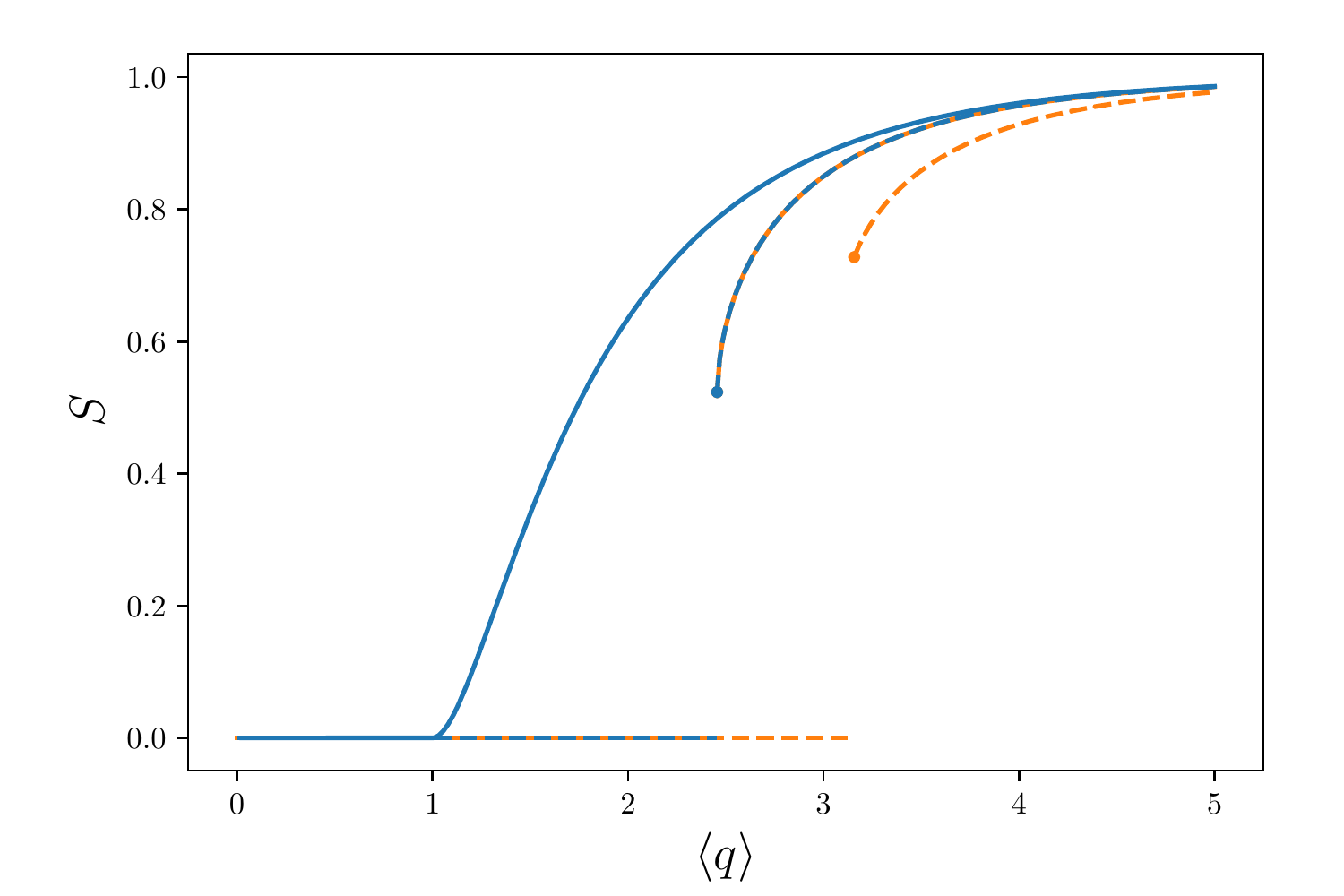}
\caption{Relative size of the giant mutually connected component (MCC) as a function of mean degree in \ER networks with $M=2$ layers (blue dashed) and $M=3$ layers (orange dashed). In both cases the transition is discontinuous. Also shown are the corresponding sizes of the weak multiplex percolation giant component (solid blue and orange lines for $M=2$ and $3$, respectively). 
The size of the giant MCC for two symmetric \ER layers coincides with the size of the weak percolation giant component in three layers.
}\label{rapidlyExamples_MCC}
\end{figure}

The mutually connected clusters in a multiplex or multi-layer interdependent network are groups of nodes, each
pair of which is connected by a path in every layer of the multiplex network
to which they both belong \citep{buldyrev2010catastrophic}. In other words, a group of nodes which belong to the same connected component in every layer in which they participate. This rule is more strict than the weak multiplex percolation, so any giant mutually connected component is a subcomponent of the giant weak percolating cluster.

\begin{boxedtext}{Identifying the mutually connected clusters}
To identify the mutually connected components, one must pay attention to the clusters to which each node belongs. One may use the following pruning algorithm \citep{baxter2012avalanche}:

\begin{enumerate}
\item Choose a test node $i$ at random from the network.
\item For each kind of edge $s$, compile a list of vertices that can be
reached from $i$ by following only edges of type $s$.
\item The intersection of these $m$ lists forms a new candidate set for
the viable cluster containing $i$.
\item Repeat steps $2.$ and $3.$ but traversing only the current candidate
set. When the candidate set no longer changes, it is either a
viable cluster, or contains only node $i$.
\item To find further viable clusters, remove the viable cluster of $i$
from the network (cutting any edges) and repeat steps $1.$ to $4.$ 
on the
remaining network beginning from a new test node.
\end{enumerate}
\end{boxedtext}

While one may invent other algorithms than the one given here to achieve the same result, the exploration of connected clusters at each step is unavoidable.
It is not possible to identify
whether a node is a member of a mutually connected cluster simply by examining its
immediate neighbours. 
Compare with the much more rapid algorithm given above for identifying the weak percolating clusters. This highlights one of the principle differences between the two rules.

The critical behaviour of this process can once again be examined through the use of self consistency equations.
As before, we consider network layers which are sparse, uncorrelated networks, in the large size limit, with the  joint degree distribution $P(q_a,q_b,...)$.
Let us define $X_s$, with  $s \in \{a,b,...\}$,
to be the probability that, upon following an
arbitrarily chosen edge of type $s$ (layer $s$), we encounter the root of an
infinite sub-tree formed solely from type $s$ edges, whose nodes
are also each connected to at least one infinite subtree of every
other type. We call this a type $s$ infinite subtree.
Writing these conditions in the form of self consistency equations, we arrive at  \citep{baxter2012avalanche}:
\begin{align}\label{Psi_s}
X_s =&
\sum_{q_a,q_b,...}\!\!\!\!\frac{q_s}{\langle q_s\rangle}
P(q_a,q_b,...) \big[1-(1-X_s)^{q_s-1} \big] \!\prod_{l\neq s}\!
\big[1-(1-X_l)^{q_l}\big]\nonumber\\
 \equiv & \Phi_s(X_a,X_b,...)\,.
\end{align}

The term
$[1-(1\!-\!X_a)^{q_a}]$ is the probability that the encountered node has at
least one edge of type $a \neq s$ leading to the root of an
infinite sub-tree of type $a$ edges. This
becomes $[1-(1\!-\!X_s)^{q_s-1}]$ when $a = s$.

Notice that $\Phi_s$ depends both on $X_s$ itself, as well as on all the other layers, while in the equivalent equation for weak multiplex percolation, Equation (\ref{Z_general}), there is no dependence on the layer we are considering. The extra condition means that the giant mutually connected component equations appears with a discontinuous hybrid transition even in two layers, as illustrated in Figure \ref{rapidlyExamples_MCC}.

The probability, $S$, that a randomly selected node belongs to the giant mutually connected component, can then be calculated in terms of these probabilities.
A node is  in the giant mutually connected cluster if it has at least
one edge of every type $s$ leading to an infinite type $s$ sub-tree
(probability $X_s$):
\begin{equation}\label{S}
S = 
\sum_{q_a,q_b,...}\!\! P(q_a,q_b,...)\!\prod_{s=a,b,...}\!\!
\big[1-(1\!-\!X_s)^{q_s} \big],
\end{equation}
This is, of course, equal to the relative size of the giant mutually connected cluster.


\section{Critical phenomena with rapidly decaying degree distributions}
\label{rapidlyDecaying}

Let us now return to weak multiplex percolation, and examine how the self consistency equations given in Section \ref{selfconsistencyWP}, above, allow us to characterise the critical phenomena of weak multiplex percolation.
If the degree distributions of each layer are sufficiently rapidly decaying, their first few moments are finite. This is the case for, for example, \ER network layers, which have Poisson degree distributions, or any other distribution with an exponentially decaying tail. In fact the following results are valid as long as the degree distribution decays faster than $q^{-3}$ for large degrees $q$.
For heavy tailed degree distributions, more exotic critical phenomena are observed, which we will explore in Section \ref{heavyTailed}.

The observed behaviour is qualitatively different in two layers than in three or more layers, so we will examine these cases separately.

\subsection{Two layers}

When there are two layers, $M=2$, the giant component emerges with a continuous phase transition, and grows quadratically above the critical points, in contrast to the ordinary (single layer) percolation transition which has linear growth.

In this case the size of the giant weak percolation cluster is given by
\begin{equation}
S
= \sum_{q_a,q_b} P(q_a,q_b) 
[1-(1-Z_a)^{q_a}][1-(1-Z_b)^{q_b}]
,
\label{270}
\end{equation}
where, from Equation (\ref{Z_general}),
\begin{eqnarray}
Z_a &=& \sum_{q_a,q_b} \frac{q_a P(q_a,q_b)}{\langle q_a \rangle} 
\,\,[1-(1-Z_b)^{q_b}]
,  
\nonumber
\\[3pt]
Z_b &=& \sum_{q_a,q_b} \frac{q_b P(q_a,q_b)}{\langle q_b \rangle} 
[1-(1-Z_a)^{q_a}]\,.
\label{260}
\end{eqnarray}

One may easily solve these equations numerically to obtain the size of the giant component for any joint degree distribution, and in particular at any network density, controlled, for example, through varying the mean degree.
We may also, however, obtain from these equations analytic results for the critical point and behaviour of the giant component near it.

Assuming a continuous transition, $S$ and hence the probabilities $Z_a$ and $Z_b$ are small close to the critical point. We therefore expand Equations (\ref{Z_general}) in powers of $Z_a$ and $Z_b$:
\begin{align}
Z_a &= \sum_{q_a,q_b} \frac{q_a P(q_a,q_b)}{\langle q_a \rangle} 
[
q_bZ_b - \frac12 q_b(q_b-1)Z_b^2 + ...
]
\nonumber\\
 &=  \frac{1}{\langle q_a \rangle} 
\bigl[
\langle q_a q_b \rangle Z_b - \frac12 \langle q_a q_b (q_b -1 ) \rangle Z_b^2
\bigr] + \mathcal{O}(Z_b^3)\,,\label{272a}
\\
Z_b &=  \frac{1}{\langle q_b \rangle} 
\bigl[
\langle q_a q_b \rangle Z_a - \frac12 \langle q_a q_b (q_a -1 ) \rangle Z_a^2
\bigr] + \mathcal{O}(Z_a^3)\,,
\label{272}
\end{align}
for $Z_a, Z_b \ll 1$, where the summations over degree can be evaluated 
 to give expressions in terms of the first few moments of the joint degree distribution $P(q_a,q_b)$, namely $\langle q_a \rangle$, $\langle q_b \rangle$, $\langle q_a q_b\rangle$, $\langle q_a^2 q_b \rangle$, and $\langle q_a q_b^2 \rangle$. %

We may identify the threshold for the continuous transition by taking the limit $Z_a,Z_b \to 0+$. 
Keeping only the linear terms in Equations (\ref{272a}) and (\ref{272}) we substitute the second equation into the first (or vice versa), finding
\begin{equation}
Z_a \cong   \frac{1}{\langle q_a \rangle \langle q_b \rangle} \langle q_a q_b \rangle^2  Z_a\,.
\end{equation}
A valid nonzero solution is only obtained if 
\begin{equation}
 { \langle q_a q_b \rangle^2}-{\langle q_a \rangle \langle q_b \rangle} = 0\,
\label{276}
\end{equation}
hence this is the condition for the critical point.

To obtain the growth of the giant component above this point, we return to Equations (\ref{272a}) and  (\ref{272}), now keeping also second order terms. Again, we substitute one into the other, Equation (\ref{272}) into (\ref{272a}), say. Keeping only up to quadratic terms, and solving for $Z_a$ we have
\begin{multline}
Z_a = 
2 \left(1 - \frac{\langle q_a q_b \rangle^2 }{\langle q_b \rangle\langle q_a \rangle}\right)
\Bigg[\frac{\langle q_a q_b \rangle \langle q_a q_b (q_a -1 ) \rangle }{\langle q_b \rangle\langle q_a \rangle^2} 
\\ +
\frac{\langle q_a q_b \rangle^2 \langle q_a q_b (q_b -1 ) \rangle }{\langle q_b \rangle^2\langle q_a \rangle}
\Bigg] ^{-1}\label{Za_growth}
\end{multline}		
and similarly for $Z_b$, which may be obtained by exchanging the indices $a$ and $b$.

Making a similar expansion of Equation (\ref{270}) gives
\begin{equation}
S
\cong \langle q_a q_b \rangle Z_a Z_b  
,  
\label{274}
\end{equation}
then simply substituting Equation (\ref{Za_growth}), and its counterpart for $Z_b$, into Equation (\ref{274}),
we find that, near the critical point,
\begin{eqnarray}
S &\cong& 
\frac{4\,{\langle q_a \rangle \langle q_b \rangle}\bigl( \langle q_a q_b \rangle^2 - \langle q_a \rangle \langle q_b \rangle \bigr)^2}
{
{\langle q_a q_b \rangle}
Q_a
Q_b
}
\label{278}
\end{eqnarray}
where for compactness we have defined
\begin{align}
Q_a \equiv \bigl[  \langle q_a q_b(q_a {-} 1) \rangle \langle q_b \rangle + \langle q_a q_b(q_b {-} 1) \rangle \langle q_ a q_b \rangle \bigr],\\
Q_b \equiv \bigl[  \langle q_a q_b(q_b {-} 1) \rangle \langle q_a \rangle + \langle q_a q_b(q_a {-} 1) \rangle \langle q_ a q_b \rangle \bigr].
\end{align}
Comparing with Equation (\ref{276}), we see that the size of $S$ is proportional to the square of the distance from the critical point. 

These results become a little simpler, and more transparent, in the less general case in which degrees in different layers are uncorrelated. That is, $P(q_a,q_b)=P_a(q_a)P_b(q_b)$.  Then $\langle q_aq_b\rangle = \langle q_a\rangle \langle q_b\rangle$, so the condition for the critical point, Equation (\ref{276}), reduces to
\begin{equation}
\langle q_a \rangle \langle q_b \rangle = 1
\label{292}
\end{equation}
and the size of the giant component near the transition becomes
\begin{eqnarray}
&& 
\!\!\!\!\!
S \cong 
\frac{4\langle q_a \rangle^3(\langle q_a \rangle\langle q_b \rangle - 1)^2}
{[(\langle q_b^2 \rangle \langle q_a \rangle^2 {-} 1)\langle q_a \rangle + \langle q_a^2 \rangle - \langle q_a \rangle^2]^2  
} 
\label{296}
\end{eqnarray}
where we have used that $\langle q_b\rangle = 1/\langle q_a\rangle$ at the critical point.
It is clear that $S$ is quadratic in the distance from the critical point in the $\langle q_a\rangle-\langle q_b\rangle$ plane.

In the symmetric case, where the degree distribution in both layers is the same ( $P_a(q_a) = P_b(q_b) \equiv P(q)$), the critical point occurs when the mean degree (common to both layers) is one: $\langle q \rangle =1$.
The size of the giant component near the transition becomes 
\begin{equation}
S
\cong 4\, \frac{(\langle q \rangle-1)^2}{(\langle q^2 \rangle-1)^2} 
.  
\label{298}
\end{equation}
One can use the mean degree $\langle q \rangle$ as a control parameter, so that the distance from the critical point is $(\langle q \rangle-1)$. One sees clearly that $S$ grows quadratically in this distance.

For two \ER network layers
the degree distribution in each layer is Poisson, $P(q) =  c^qe^{-c}/q!$ where $c$ is the mean degree. In this specific case, the self consistency equations take a very convenient form.
Equation (\ref{270}) becomes
\begin{align}
S =(1-  e^{-c_a Z_a})
 (1- e^{-c_b Z_b})
\end{align}
while Equation (\ref{260}) becomes
\begin{align}
Z_a &= 1- e^{-c_bZ_b}\,,
\nonumber
\\
Z_b 
&= 1- e^{-c_aZ_a}\,.
\end{align}

Using the fact that the second moment of a Poisson distribution is $c^2+c$, and that at the critical point $c_ac_b = 1$, see Equation (\ref{292}), Equation (\ref{296}) becomes
\begin{align}
S &\cong 
\frac{4(c_ac_b - 1)^2} {(c_b+1) (c_a+1) } 
\end{align}
for $c_a$, $c_b$ close to $1$.

The solution for two symmetric \ER network layers (that is, with the same mean degree) is plotted in Figure \ref{rapidlyExamples}.
In this case, Equation (\ref{298}) becomes
\begin{equation}
S \cong \frac{4(c-1)^2}{(c^2+c-1)^2} 
.  
\end{equation}


\subsection{Three or more layers}
\label{s5} 

When there are more than two layers, the giant weak percolation cluster appears with a discontinuous, hybrid transition  \citep{baxter2014weak,baxter2016unified}. At the critical point, the giant cluster jumps in size from zero to a finite fraction of the network, and then grows as the square root of the distance above the critical point. This discontinuity and the square root singularity are associated with avalanches whose mean size diverges in size as the critical point is approached from above. The mean size of avalanches plays the role of susceptibility, hence the transition has some features in common with both second- and first-order phase transitions. \
Avalanches occur when the failure of a node removes an essential supporting connection of a neighbour, which in turn fails, and so on. As we show below, one may map directed clusters of nodes in such a critical state, and thus calculate the mean size of avalanches. This diverges as the inverse square root of the distance above the critical point.

This type of avalanche process is responsible for hybrid transitions of the same type observed in the mutually connected cluster \citep{baxter2012avalanche}, in bootstrap percolation \citep{baxter2010bootstrap}  and in $k$-core percolation \citep{dgm2006,baxter2015critical}. The difference between two and more than two layers in weak multiplex percolation is analogous to the difference between threshold $k=2$ and $k>2$ in $k$-core percolation \citep{dgm2006}.

The discontinuous hybrid transition occurs at the point where a second solution to Equation (\ref{Z_general}) appears, at a point where the surface defined by $\bf{\Psi}(\bf{Z})$ is tangent to the plane $\bf{Z}$. This happens when 
\begin{align}
\det[\mathbf{J} - \mathbf{I}] = 0
\end{align}
where $\mathbf{J}$ is the Jacobian matrix $J_{lm} = \partial\Psi_l /\partial Z_m$, and $\mathbf{I}$ is the identity matrix \citep{baxter2014weak, baxter2016unified}.


To gain some intuition about this behaviour, in $M>2$ layers, let us first consider the symmetric, uncorrelated case 
\begin{equation}
P(q_1,q_2,...,q_M) = \prod_{l = 1}^M P(q_l)\,.
\end{equation}
Then Equations (\ref{Z_general}) become a single equation
\begin{align}
Z &= 
 [1-\sum_{q}
P(q)(1-Z)^{q}]^{M-1}\nonumber\\
&=  [1 - G(1-Z)]^{M-1}
 \equiv \Psi(Z), \label{580}
\end{align}
where $G(x)$ is the degree distribution generating function
\begin{equation}\label{Gdefn}
G(x) = \sum_{q} P(q) x^q\,.
\end{equation}
The relative size of the giant component is then given by 
\begin{align}
S = [1 - \sum_{q} P(q)(1-Z)^{q}]^M =  [1 - G(1-Z)]^{M} = Z^{M/(M-1)}\,.\label{590}
\end{align}

A hybrid transition occurs when the line $Z$ is tangent to $\Psi(Z)$, which occurs when
\begin{equation}\label{hybrid_crit}
\left(\frac{\Psi}{Z}\right)' = \frac{1}{Z}\left[ \Psi' - \frac{\Psi}{Z}\right] = 0\,.
\end{equation}
Substituting in $\Psi(Z)$ from Equation (\ref{580}) one quickly finds the condition for the critical point
\begin{align}
Z_c = \left[(M-1) G'(1-Z_c)\right]^{-(M-1)/(M-2)}\,,\label{Zc_symmetric}
\end{align}
hence
\begin{align}
S_c = \left[(M-1) G'(1-Z_c)\right]^{-M/(M-2)}\,.\label{Sc_symmetric}
\end{align}
where the subscript $c$ indicates that this is the value at the critical point.
Solving Equation (\ref{Zc_symmetric}) for $Z_c$, and substituting back into Equation (\ref{580}) one can find the value of the control parameter at the point of the  hybrid transition.

To find the scaling near the critical point, we expand 
Equation (\ref{580}) about the critical value $Z_c$, given by Equation (\ref{Zc_symmetric}).
Note that the function $\Psi$ depends also on the control parameter $\langle q \rangle$ through the generating function $G$.
We therefore expand $\Psi$ in both $\langle q\rangle - \langle q\rangle_c$, to first order and $Z-Z_c$, to second order:
\begin{align}
Z  &= \Psi(Z,\langle q\rangle)\nonumber\\
&= \Psi(Z_c,\langle q\rangle_c) 
+ (\langle q\rangle - \langle q\rangle_c) \partial_{\langle q\rangle}\Psi(Z_c,\langle q\rangle_c) 
\nonumber\\
&\qquad + (Z-Z_c)  \partial_{Z}\Psi(Z_c,\langle q\rangle_c) 
+ \frac12 (Z-Z_c)^2 \partial_{ZZ}\Psi(Z_c,\langle q\rangle_c) \,.
\end{align}
Now, $\Psi(Z_c,\langle q\rangle_c)  = Z_c$, and the condition for the critical point, Equation (\ref{hybrid_crit}), means that the first derivative of $\Psi $ with respect to $Z$ is equal to one at the critical point, thus these two terms cancel with $Z$ on the left hand side, giving
\begin{align}
 (\langle q\rangle - \langle q\rangle_c) \partial_{\langle q\rangle}\Psi(Z_c,\langle q\rangle_c) 
&=  -\frac12 (Z-Z_c)^2 \partial_{ZZ}\Psi(Z_c,\langle q\rangle_c) \,.
\end{align}
We can therefore conclude that
\begin{equation}\label{squareroot}
Z - Z_c \propto [\langle q\rangle-\langle q\rangle_c]^{1/2}.
\end{equation}
Since $S -S_c \propto Z -Z_c$ when we are close to the threshold, the size of the giant component exhibits the same square root singularity:
\begin{equation}
S - S_c \propto [\langle q\rangle-\langle q\rangle_c]^{1/2}.
\end{equation}
This square-root scaling is the typical behaviour of the order
parameter near a hybrid transition. It results from avalanches of
spreading damage which diverge in size near the transition, see Section \ref{avalanches} below.

For the example of identically distributed \ER network layers, the degree distribution is Poisson, and one can use the mean degree $c$ as control parameter. 
In this case the generating function takes the simple form $G(1-Z) = e^{-c Z}$, and its derivative $G'(1-Z) = c e^{-c Z}$.
Equation( \ref{580}) then becomes
\begin{equation}
Z = [1 - e^{-cZ}]^{M-1}\,,
\label{592}
\end{equation}
and Equation (\ref{590}) becomes
\begin{align}
S = [1 -  e^{-cZ} ]^M = Z^{M/(M-1)}\,.\label{590}
\end{align}
One may easily obtain the solution for $Z$, and hence $S$, numerically.
The solution for $M=3$ \ER layers, all with the same mean degree, is plotted in Figure \ref{rapidlyExamples}.

The criterion for the critical point becomes
\begin{align}
\frac1Z_c = e^{-cZ_c} \left[(M-1)c+\frac1Z_c\right]
\end{align}
subject to $Z_c$ simultaneously being a solution to Equation (\ref{592}).


The simple form of the equations in the case of \ER layers reveals an interesting aspect of the relationship between weak multiplex percolation and the mutually connected component.
Equation (\ref{592}) is practically identical to the one obtained in Ref.~\citep{gao2011robustness} for the relative size $S^\ast$ of the giant mutually connected component in $M$-layer multiplex \ER networks: 
\begin{equation}
S^\ast = [1 - e^{-cS^\ast}]^M
\label{594}
.
\end{equation}
Comparing Eqs.~(\ref{590}) and (\ref{592}) with Equation ~(\ref{594}), we obtain the following relation between these two problems, for a given mean degree $c$:
\begin{align}
S(c,M) &= S^{\ast\,M/(M-1)}(c,M-1).
\end{align}
where $M \geq 3$.
In terms of the critical value of the mean degree, $c_c$, the relation is even closer:
\begin{align}
c_\text{c}(M) &= c^\ast_\text{c}(M-1)
,
\label{596}
\end{align}
where $c_c(M)$ and $c_c^\ast(M)$ are the critical value of the average degree for the giant mutually connected component and weak percolation, respectively, for $M$-layer multiplex \ER networks. 
Thus the weak percolation problem on an $M$-layer multiplex \ER network is equivalent to the problem of giant mutually connected component in the corresponding $M-1$-layer multiplex \ER network. This can be seen clearly in Figure \ref{rapidlyExamples_MCC}, showing that the solution for weak percolation in $M=3$ layers is identical to the solution for the mutually connected component in $M=2$ layers. 

For large $M$, the asymptotics of these quantities are the same in both problems \citep{baxter2020weak}: 
\begin{eqnarray}
c_\text{c} &\cong& \ln M + \ln\ln M + 1 + \frac{\ln\ln M}{\ln M}
,
\nonumber
\\[3pt]
S_\text{c} &\cong& 1 - \frac{1}{\ln M} + \frac{\ln\ln M}{\ln^2 M}
.
\label{598}
\end{eqnarray}



\subsubsection{Avalanches}\label{avalanches}

To understand the discontinuous appearance of the giant weak percolating cluster, we now describe how the statistics of cascading avalanches may be calculated.
Let us define  a critical node of type $l$ to be one that only just meets the criteria for
inclusion in the weak percolating cluster. This occurs when a node has one and only one connection to an active node in layer $l$, and at least one in all the other layers to which it belongs. We call this connection a critical edge of type $l$.
A  node
may be critical with respect to more than one layer, if it
simultaneously has exactly one connection to active nodes in several layers.

\begin{figure}[htb]
\begin{center}
	\includegraphics[width=0.6\columnwidth]{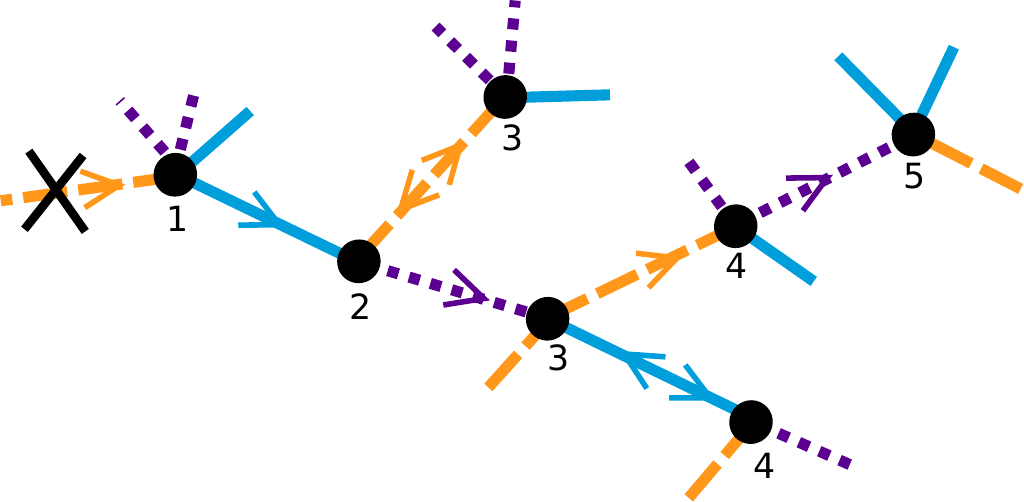}
	\caption{A small group of critical nodes in a three layer multiplex network. Edges in different layers are shown in different colours and with different line patterns. Open ended edges are considered to lead to active nodes which are not in a critical state. Critical edges and the direction of dependency are indicated by arrows. 
	If such an edge is removed, the node it leads to will fail to meet the criteria for inclusion in the weak percolating cluster, and become inactive. This may lead to the removal of further critical edges and hence the removal of further nodes. Note that some critical edges have arrows in both directions.
	For example, if the edge marked with a cross at the left of the figure will cause the node marked $1$ to become inactive. This leads to the failure of the node marked $2$, then those marked $3$, $4$ and $5$ in turn. This is the process by which avalanches of damage propagate through the multiplex network.
	}\label{critical_cluster}
\end{center}
\end{figure}

If such a node loses a critical edge, it will become inactive. If any of its other (outgoing) edges, in any layer, happen to be critical edges of other critical nodes, these in turn will be removed from the weak percolating cluster. In this way, damage to the network may propagate through the multiplex, an avalanche.
An example is shown in Figure \ref{critical_cluster}.
Note that we may think of critical edges as being directed, in that the removal of such an edge will provoke the removal of the node at one of its ends, but not necessarily the other. The path of an avalanche through the network may therefore be traced by following the directed critical edges.

Let us consider the possible configurations we encounter when following a randomly chosen edge in layer $l$:

i ) With probability $1-Z_l$ the node we encounter fails to have any connection to active nodes in at least one layer.

ii) With probability $Y_l$ the node we encounter has at least one connection (not counting the edge along which we arrived) to active nodes in all layers (probability $Z_s$).

iii) With probability $R_l$ the node encountered has no further connections to active nodes in layer $l$, and at least one connection (not counting the edge along which we arrived) to active nodes in all other layers.

The probability $Z_l$ is the sum of these two probabilities, $Z_l = Y_l + R_l$. We distinguish these cases, because $R_l$ gives the probability that the edge we followed is a critical edge of type $l$ for the node we arrive at, and can therefore be used to calculate avalanche statistics.
We can write the following self-consistency equation for $R_l$:
\begin{equation}\label{eq:R1_WPP}
R_l = \sum_{q_1,q_2,...,q_M}\frac{q_l P(q_1,q_2,...,q_M)}{\langle q_l\rangle}
(1-Z_l)^{q_l-1} \prod_{n\neq l}
\left[ 1 - (1-Z_n)^{q_n}\right].
\end{equation}

The critical node reached may have further critical nodes emanating from it in other layers (which occurs with probability $R_s$ in layer $s$), leading to further critical nodes. The subtree formed by including only such critical edges and the critical nodes that they lead to gives the size of the avalanche provoked by removing the original edge (for example, if the node it originates from becomes inactive).
We may calculate the statistics of such subtrees using generating functions.
Let us define $H_l(\bf{u})$ to be the multivariate generating function for the size of the critical subtree encountered upon following an edge in layer $l$. 
Where ${\bf u}$ is a vector of $M$ variables, $u_l$, one for each layer. 
The generating function should be a sum over all possible sizes for the critical subtree, including a factor $u_l$ for every critical node reached along an edge of type $l$.
This generating function may be defined in a recursive way,
\begin{equation}\label{eq:H1_WPP}
H_l({\bf u}) = (Z_l - R_l) + u_l F_l[H_1({\bf u}), H_2({\bf u}),...,H_M({\bf u})].
\end{equation}
Where the functions $F_l({\bf x})$ are given by
\begin{multline}\label{eq:F1_WPP}
F_l({\bf x}) =  \sum_{q_1,q_2,...}\frac{q_l P(q_1,q_2,...)}{\langle q_l\rangle}
(1 - Z_l)^{q_l-1} \prod_{n\neq l}
\sum_{s=1}^{q_n} \binom{q_n}{s} (1 - Z_n)^{q_n-s} x_n^{s}.
\end{multline}

The first term in Equation (\ref{eq:H1_WPP}) simply gives the probability $Y_l$ that the edge is not critical, so the subtree has size zero.
The second term includes one factor of $u_l$, times a probability given by the function $F_l$. Compare the form of Equation (\ref{eq:F1_WPP}) with Equation (\ref{eq:R1_WPP}). The function calculates the probability that an edge in layer $l$ satisfies the condition for a critical edge, as well as including a factor $x_n$ for each outgoing edge from the node reached, in each layer $n \neq l$. For each of these edges, we may apply the generating function again, replacing $x_n$ with the corresponding generating function $H_n({\bf u})$. Notice that $F_l$ has no dependence on $x_l$.
This method is very similar to that used in \citet{baxter2012avalanche}.
The mean size of the avalanche caused by the removal of single node
is then given by 
\begin{equation}
\sum_{l} \partial_{u_l}H_l({\bf 1})\,.
\end{equation}
Where $\partial_z$ signifies the partial derivative with respect to
variable $z$.
Note that $F_l(Z_1,Z_2,...,Z_M) = R_l$ and also that
$H_l(1,1,...,1) = Z_l$.

To demonstrate the behaviour of avalanches, we return again to the symmetric case, in which all layers are random graphs constructed according to the same degree distribution $P(q)$.
In this case, there is only a single variable $Z$, a single $R$, and a single generating function $H(u)$. Equation (\ref{eq:R1_WPP}) becomes 
\begin{equation}\label{eq:R1_WPPsym}
R = \left[1 - \sum_q P(q) (1-Z)^q \right]^{M-1} \sum_q \frac{qP(q)}{\langle q\rangle}(1-Z)^{q-1}
\end{equation}
while Equations (\ref{eq:H1_WPP}) and (\ref{eq:F1_WPP}) become
\begin{equation}\label{eq:H1_WPPsym}
H(u) = (Z - R) + u F[H(u)].
\end{equation}
and
\begin{align}
F({x}) &=  \left\{
\sum_q P(q) \sum_{s=1}^q \binom{q}{s} (1-Z)^{q-s} x^s
\right\} ^{M-1}\sum_q \frac{qP(q)}{\langle q\rangle}(1-Z)^{q-1}\nonumber\\
&= 
\left\{
\sum_q P(q)
\left[(1-Z+x)^q - (1-Z)^q\right]
\right\} ^{M-1}\sum_q \frac{qP(q)}{\langle q\rangle}(1-Z)^{q-1}. \label{eq:F1_WPPsym}
\end{align}
respectively.

The mean avalanche size is then given by $H'(1)$.
	\begin{align}
	H'(1) &=  
	R + u F'[Z] H'(1) 
	\\
	\Rightarrow H'(1) &= \frac{R}{ 1 - F'(Z)}
	.
	\end{align}
Now, in this case
\begin{align}
F'(Z) &= 
(M-1) 
\left[ 1 -\sum_q P(q) (1-Z)^q\right]
^{M-2}
\sum_q 
{qP(q)}
(1-Z)^{q-1}.
\end{align}
If we compare with the derivative of the right hand side of Equation (\ref{580}), we see that $F'(Z) = \Psi'(Z)$.
Thus
\begin{align}
H'(1) = \frac{R}{ 1 - \Psi'(Z)}\,.
\end{align}
Noting that at the critical point $\Psi'(Z) = 1$, since $\Psi(Z)$ is tangent to $Z$, we see that the mean avalanche size diverges at this critical point.
These diverging avalanches are responsible for the discontinuity in the size of the giant weak percolation component at the critical point.

\section{Critical phenomena with broad degree distributions}\label{heavyTailed}

For strongly heterogeneous degree distributions the condition that the leading moments are finite may not be met. 
In this case we may use generating functions to study the asymptotics of the solutions. 

For concreteness, we will consider uncorrelated powerlaw degree distributions of the form 
\begin{align}\label{powerlaw}
P(q) = Aq^{-\gamma}\
\end{align}
for each layer, with possibly different values of $\gamma$ in each layer. The $m$-th moment of this distribution $\langle q^m\rangle$ diverges if $m \geq \gamma-1$. This means that the approach used in the previous Section is no longer valid, as the relevant moments diverge if $\gamma < 3$.

We may rewrite Equations (\ref{Z_general}) and (\ref{S_general}) in terms of generating functions of the form
\begin{align}
G(x) = \sum_{q} P(q) x^{q}\,.
\end{align}
One may expand $G(1-Z)$ for small $Z$, finding a term with non-integer exponent $\gamma-1$, which may be the leading, second or higher order term depending on the value of $\gamma$:
\begin{align}
G(1-x) &\cong 1 - \langle q \rangle x + A \Gamma(1-\gamma) x^{\gamma-1}+ \mathcal{O}(x^2)
\label{Gexpansion}
\end{align}
for $\gamma > 2$. If $1 < \gamma < 2$, the coefficient of the linear term is not the mean degree (which diverges) but a constant $B$ which depends on the specifics of the degree distribution, 
\begin{align}
G(1-x) &\cong 1 + Bx + A \Gamma(1-\gamma) x^{\gamma-1}+ \mathcal{O}(x^2)\,.
\label{Gexpansion2}
\end{align}
See Appendix \ref{sa2} for details.

\subsection{Two Layers}

In two layer multiplex networks with powerlaw degree-distributed layers, the critical point remains at $\langle q \rangle = 1$ for $\gamma>2$, but  the 
exponent of $2$ for the growth of the giant component above the critical point, ($\beta$ exponent) does not apply for $\gamma < 3$. Instead, we find that it depends on the value of $\gamma$. Compare this with the case of ordinary percolation, for which the standard exponent $\beta=1$ applies only for $\gamma > 4$.

With uncorrelated layers,  Equations~(\ref{260}) can be rewritten in terms of generating functions (see Appendix \ref{sa2}) as 
\begin{align}
Z_a &= 1 - G_b(1-Z_b)
,  
\nonumber \\
Z_b &= 1 - G_a(1-Z_a) 
,
\label{300}
\end{align}
and the size of the giant percolating cluster is then simply
\begin{equation}
S = Z_aZ_b   
.
\label{301}
\end{equation}


\begin{figure}
\centering
\includegraphics[width=0.6\textwidth]{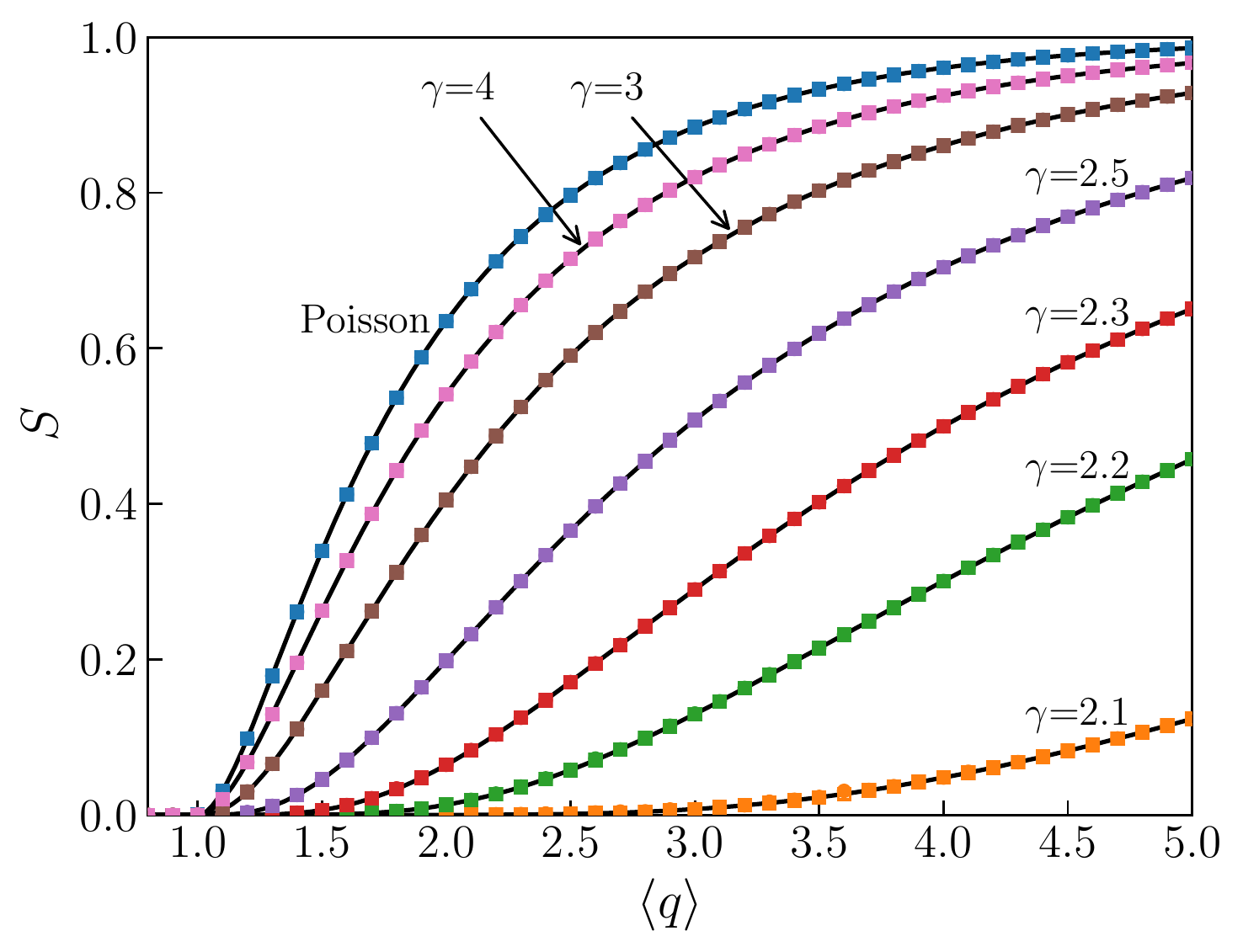}
\caption{Relative size $S$ of the weak multiplex percolation giant component as a function of mean degree $\langle q\rangle$ in two layer multiplex network, $M = 2$, with each layer having the same powerlaw degree distribution $P(q) \sim Aq^{-\gamma}$. Theoretical curves (solid lines) are compared with simulation results for the relative size of the largest cluster for networks containing  $N=10^6$ and $N=10^7$ Results for Poisson degree distributions are shown for comparison\protect\footnotemark.
}\label{fig_M2}
\end{figure}
%
\footnotetext{Figure reproduced from \citep{baxter2020weak} with permission of the authors.}

Let us first consider the illustrative case of symmetric layers, $P(q_a) = P(q_b) \equiv P(q)$.
When $2<\gamma<3$, using Equation (\ref{Gexpansion}), 
\begin{equation}
Z \cong \langle q \rangle Z - A\Gamma(1-\gamma)Z^{\gamma-1}
,  
\label{380}
\end{equation}
which is easily solved for $Z$, so that the size of the giant component is
\begin{equation}
S = Z^2 \cong \Bigl[\frac{\langle q \rangle-1}{A\Gamma(1-\gamma)}\Bigr]^{2/(\gamma-2)}
.  
\label{400}
\end{equation}
We see that the giant component grows with an exponent $2/(\gamma-2)$ above the critical point $\langle q \rangle = 1$. This exponent becomes very large close to $\gamma=2$, corresponding to very slow growth above the critical point, as can be seen in Figure \ref{fig_M2}. As $\gamma$ approaches $3$ we return to the limiting value of the exponent $\beta=2$.

These theoretical results are in excellent agreement with numerical simulations, as shown in Figure \ref{fig_M2}. Simulation results are reproduced here from \citep{baxter2020weak}. In this range of $\gamma$, simulation results converge rapidly to theory as the system size increases.


When $1<\gamma<2$, the mean degree diverges. The self consistency equations are not guaranteed to hold. Surprisingly, we find that they still give excellent results, confirmed by numerical simulations \citep{baxter2020weak}.
Since the mean degree diverges, we must use some other control parameter. 
Here we apply random damage to the multiplex network, removing a fraction $1-p$ of all edges. 
The fraction $p$ of surviving 
edges is then the control parameter.
The tail of the degree distribution retains the same powerlaw exponent $\gamma$, with a reduced amplitude \begin{align}
A_p = A_1p^{\gamma-1},\label{Ap}
\end{align}
where $A_p$ is the resulting amplitude with control parameter $p$, and $A_1$ is the undamaged amplitude ($p=1$). See Appendix \ref{sa1}.
Substituting the expansion Equation (\ref{Gexpansion2}) into Equations (\ref{300}) and (\ref{301}), in this range of $\gamma$ we find
\begin{equation}
Z \cong  [- A_p\Gamma(1-\gamma)]Z^{\gamma-1} = [- A_1\Gamma(1-\gamma)]p^{\gamma-1}Z^{\gamma-1}
,  
\label{410b}
\end{equation}
whose solution is
\begin{align}
Z = [- A_1\Gamma(1-\gamma)]^{1/(2-\gamma)}p^{(\gamma-1)/(2-\gamma)}
\end{align}
and so 
\begin{equation}
S = Z^2 \cong [-A_1\Gamma(1-\gamma)]^{2/(2-\gamma)} p^{(2(\gamma-1)/(2-\gamma)} 
.  
\label{430b}
\end{equation}
This indicates that the giant component appears immediately from $p=0$, and grows with exponent $2(\gamma-1)/(2-\gamma)$, which is positive for $1 < \gamma < 2$.  This exponent diverges as we approach $\gamma=2$ from below.

%

\subsubsection{Non-symmetric layers}

One may carry out a similar analysis for the more general case that the degree distributions of the two layers have powerlaw tails with different decay exponents. Generally the smaller of the two exponents dictates the critical behaviour, except when both are less than $2$, see \citep{baxter2020weak}.
For concreteness, let us consider that each layer has a powerlaw degree distribution, with a different exponent for each layer:
$P_a(q_a) = Aq_a^{-\gamma_a}$ and $P_b(q_b) = Aq_b^{-\gamma_b}$.

If both exponents $\gamma$ are greater than $3$, we have the behaviour described in Section \ref{rapidlyDecaying} for rapidly decaying degree distributions. We thus consider various cases for which one or both exponents are less than three.
We assume, without loss of generality, that $\gamma_a > \gamma_b$. Results for the opposite case can be obtained by simply exchanging the subscripts $a$ and $b$. 

We may substitute the expansions of the generating function $G$, Equations (\ref{a60})-(\ref{a80}), into Equation (\ref{300}) to obtain leading order approximations for the self-consistency equations for $Z_a$ and $Z_b$ for the different possible combinations of the exponents $\gamma_a$ and $\gamma_b$.
If $\gamma_b>3$,
\begin{equation}
Z_a \cong   \langle q_b \rangle Z_b - \frac{1}{2} \langle q_b(q_b-1) \rangle Z_b^2.
\label{Z60}
\end{equation}
If $2<\gamma_b<3$, then
\begin{equation}
Z_a \cong  \langle q_b \rangle Z_b - A \Gamma(1-\gamma_b) Z_b^{\gamma_b-1}
.
\label{Z70}
\end{equation}
Note that $\Gamma(1-\gamma_b)$ is positive in this range.
If $1<\gamma_b<2$, 
\begin{equation}
Z_a \cong - A \Gamma(1-\gamma_b) Z_b^{\gamma_b-1} + BZ_b
\label{Z80}
\end{equation}
where now $\Gamma(1-\gamma_b)$ is negative.
Similar expressions are obtained for $Z_b$ (which depends on $Z_a$ and $\gamma_a$) by exchanging the subscripts $a$ and $b$.

Let us now consider each of the possible cases for $\gamma_a$ and $\gamma_b$.

\begin{itemize}
	
	\item{\underline{$\gamma_a>3$ and $2<\gamma_b<3$}}

In this case the leading terms in the expansion of $Z_b$ are linear and quadratic. We may neglect the quadratic term, so we have
\begin{eqnarray}
Z_a &\cong& \langle q_b \rangle Z_b - A_b\Gamma(1-\gamma_b)Z_b^{\gamma_b-1}
,  
\nonumber
\\[3pt]
Z_b &\cong& \langle q_a \rangle Z_a  
.
\label{440}
\end{eqnarray}
whose solution is 
\begin{equation}
Z_a  \cong \left[\frac{\langle q_a \rangle\langle q_b \rangle - 1}{A_b\Gamma(1-\gamma_b)}\right]^{1/(\gamma_b-2)} \langle q_a \rangle^{-(\gamma_b-1)/(\gamma_b-2)}  
,
\label{450}
\end{equation}
so 
\begin{equation}
S = \left[\frac{\langle q_a \rangle\langle q_b \rangle - 1}{A_b\Gamma(1-\gamma_b)}\right]^{2/(\gamma_b-2)} \langle q_a \rangle^{-\gamma_b/(\gamma_b-2)} 
.  
\label{460}
\end{equation}

\item{\underline{2<$\gamma_a<3$ and $2<\gamma_b<3$}}

If both exponents are less than three, we use Equation (\ref{Z70}):
\begin{align}
Z_a &\cong \langle q_b \rangle Z_b - A_b\Gamma(1-\gamma_b)Z_b^{\gamma_b-1}
,  
\nonumber
\\
Z_b &\cong \langle q_a \rangle Z_a - A_a\Gamma(1-\gamma_a)Z_a^{\gamma_a-1}  
.
\label{432}
\end{align}
Substituting the second line into the first gives
\begin{multline}
(\langle q_a \rangle\langle q_b \rangle - 1)Z_a 
\\
\cong 
\langle q_b \rangle A_a \Gamma(1{-}\gamma_a) Z_a^{\gamma_a-1} 
+
\langle q_a \rangle^{\gamma_b-1} A_b\Gamma(1{-}\gamma_b)Z_a^{\gamma_b-1}.
\label{433}
\end{multline}
As we have assumed $\gamma_a>\gamma_b$, the exponent $\gamma_a-1$ is larger than $\gamma_b-1$, so the first term on the right-hand side of Equation ~(\ref{433}) can be neglected, and we obtain
\begin{equation}
Z_a \cong \left[\frac{\langle q_a \rangle\langle q_b \rangle - 1}{\langle q_a \rangle^{\gamma_b-1} A_b \Gamma(1{-}\gamma_b)}\right]^{1/(\gamma_b-2)} 
,
\label{436}
\end{equation}
thus
\begin{equation}
S \cong 
\langle q_a \rangle^{-\gamma_b/(\gamma_b-2)} \left[\frac{\langle q_a \rangle\langle q_b \rangle - 1}{ A_b \Gamma(1{-}\gamma_b)}\right]^{2/(\gamma_b-2)}
.  
\label{437}
\end{equation}

\item{\underline{$1<\gamma_b<2$, $\gamma_a > 2$}}

In this case the leading term in the equation for $Z_b$ is linear, regardless of whether $2<\gamma_a<3$ or $\gamma_a>3$. Since we won't require higher order terms, we can treat these two cases together. We have
\begin{align}
Z_a &\cong - A_b\Gamma(1-\gamma_b)Z_b^{\gamma_b-1}
,  
\nonumber
\\
Z_b &\cong \langle q_a \rangle Z_a  
.
\label{470}
\end{align}

As we saw previously, we can no longer use the mean degree as a control parameter, as the mean degree in layer $b$ diverges. We instead
 apply random damage, using the edge occupation probability $p$ as the control parameter. 
The degree distribution for layer $b$ is then asymptotically $A_{b,1} p^{\gamma_b-1}q^{-\gamma_b}$ where $A_ {b,1}\equiv A_b(p{=}1)$, i.e. the value of $A_b$ before damage is applied.
The solution to Equations (\ref{470}) is 
\begin{align}
Z_a  &{\cong} [-A_b\Gamma(1-\gamma_b)]^{1/(2-\gamma_b)} \langle q_a \rangle^{(\gamma_b-1)/(2-\gamma_b)}  
\nonumber\\
&{\cong} [-A_{b,1}\Gamma(1-\gamma_b)]^{1/(2-\gamma_b)} \langle q_a \rangle^{(\gamma_b-1)/(2-\gamma_b)} 
p^{(\gamma_b-1)/(2-\gamma_b)}
,
\label{480}
\end{align}
and hence
\begin{align}
S &= \langle q_a \rangle Z_a^2 
\cong
[-A_{b,1}\Gamma(1{-}\gamma_b)]^{2/(2-\gamma_b)} \langle q_a \rangle^{\gamma_b/(2-\gamma_b)} p^{2(\gamma_b-1)/(2-\gamma_b)} 
.  
\label{490}
\end{align}

Under node removal, Equation (\ref{480}) becomes, instead
\begin{align}
Z_a  
&{\cong} [-A_{b,1}\Gamma(1-\gamma_b)]^{1/(2-\gamma_b)} \langle q_a \rangle^{(\gamma_b-1)/(2-\gamma_b)} 
p^{\gamma_b/(2-\gamma_b)}
,
\label{480b}
\end{align}
and hence
\begin{align}
S &= 
\cong
[-A_{b,1}\Gamma(1{-}\gamma_b)]^{2/(2-\gamma_b)} \langle q_a \rangle^{\gamma_b/(2-\gamma_b)} p^{2\gamma_b/(2-\gamma_b)} 
.  
\label{490b}
\end{align}

\medskip

\item \underline{$1<\gamma_a < 2$ and $1 < \gamma_b <2$}

In all cases so far, the critical behaviour depends on the smaller of the degree distribution exponents.
Finally, when both exponents are small the critical behaviour depends on both of them.
The equations for $Z_a$ and $Z_b$ both have the form given by Equation (\ref{Z80}):
\begin{eqnarray}
Z_a &\cong& - A_b\Gamma(1-\gamma_b)Z_b^{\gamma_b-1}
,  
\nonumber\\
Z_b &\cong& - A_a\Gamma(1-\gamma_a)Z_a^{\gamma_a-1}.
\label{492}
\end{eqnarray}
The solution is easily obtained by substituting one of these equations into the other.
This leads to 
\begin{align}
S =& Z_aZ_b \nonumber\\
\cong & \Bigl[- A_b\Gamma(1{-}\gamma_b)\Bigr]^{\gamma_a/[1-(\gamma_b-1)(\gamma_a-1)]}
\Bigl[- A_a\Gamma(1{-}\gamma_a)\Bigr]^{\gamma_b/[1-(\gamma_b-1)(\gamma_a-1)]}
\label{492}
\end{align}

In the case of random edge removal, where  the fractions of retained edges in layers $a$ and $b$ are $p_a$ and $p_b$ respectively, this is
\begin{multline}
S \cong 
\Bigl[- A_{a,1}\Gamma(1{-}\gamma_a)\Bigr]^{\gamma_b/[1-(\gamma_b-1)(\gamma_a-1)]}
\Bigl[- A_{b,1}\Gamma(1{-}\gamma_b)\Bigr]^{\gamma_a/[1-(\gamma_b-1)(\gamma_a-1)]} \\
\times p_a^{(\gamma_a-1)\gamma_b/[1-(\gamma_b-1)(\gamma_a-1)]} 
p_b^{(\gamma_b-1)\gamma_a/[1-(\gamma_b-1)(\gamma_a-1)]}\,.
\label{492a}
\end{multline}

If instead we wish to consider random node removal,  for which nodes in both layers survive with probability $p$,  we must add a factor of $p$ to Equation  (\ref{492}) [compare Equation  (\ref{410b})]. Solving for $Z_a$ and $Z_b$ then substituting into $S = p Z_aZ_b$ we find
\begin{multline}
S = \cong \Bigl[- A_{a1}\Gamma(1{-}\gamma_a)\Bigr]^{\gamma_b/[1-(\gamma_b-1)(\gamma_a-1)]}
\Bigl[- A_{b1}\Gamma(1{-}\gamma_b)\Bigr]^{\gamma_a/[1-(\gamma_b-1)(\gamma_a-1)]} \\
\times 
p^{(2\gamma_a+2\gamma_b-\gamma_a\gamma_b)/[1-(\gamma_b-1)(\gamma_a-1)]}
.
\label{492b}
\end{multline}

\end{itemize}


\subsection{More than two layers}

\begin{figure}
\centering
\includegraphics[width=0.6\textwidth]{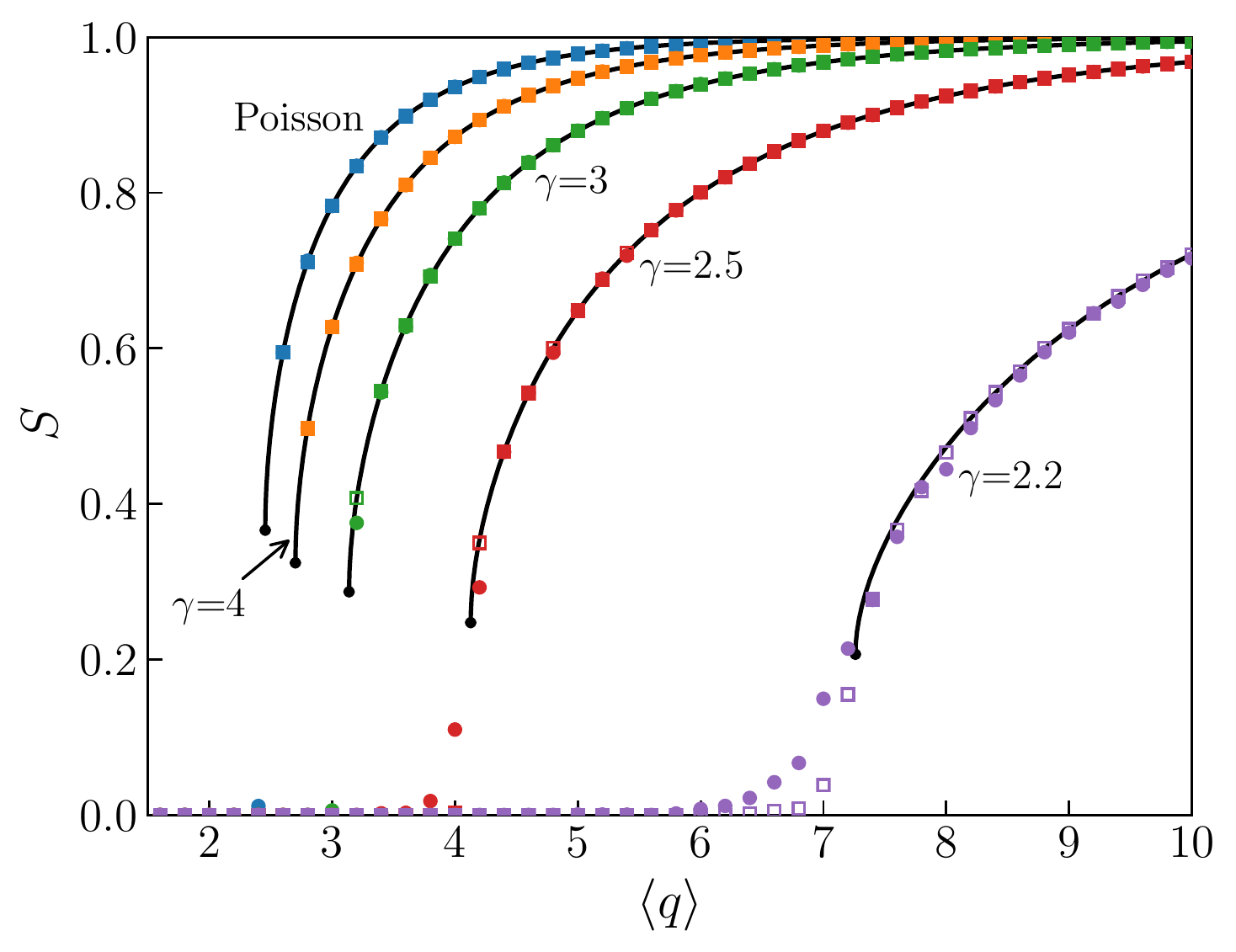}
\caption{Relative size $S$ of the weak multiplex percolation giant component for three identically powerlaw distributed layers ($M=3$) as a function of mean degree $\langle q\rangle$ for various values of $\gamma$  greater than $2$ (solid curves). Simulation results averaged over 100 synthetic networks of $N=10^4$ nodes (circles) and $N=10^7$ nodes (squares) are shown for comparison\protect\footnotemark.
\label{M3_large}}
\end{figure}
\footnotetext{Figure reproduced from \citep{baxter2020weak} with permission of the authors.}

\begin{figure}
\centering
\includegraphics[width=0.49\textwidth]{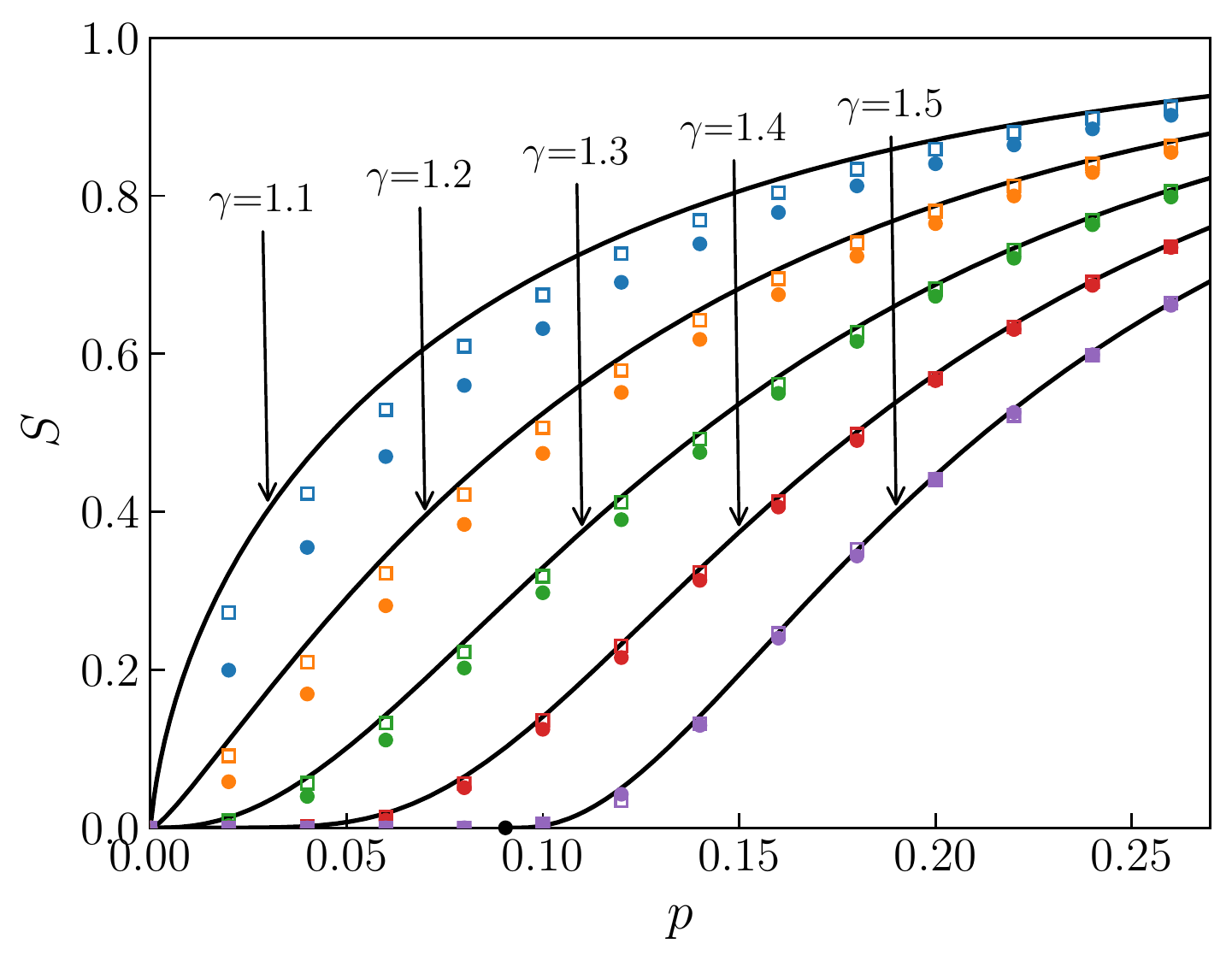}
\includegraphics[width=0.5\textwidth]{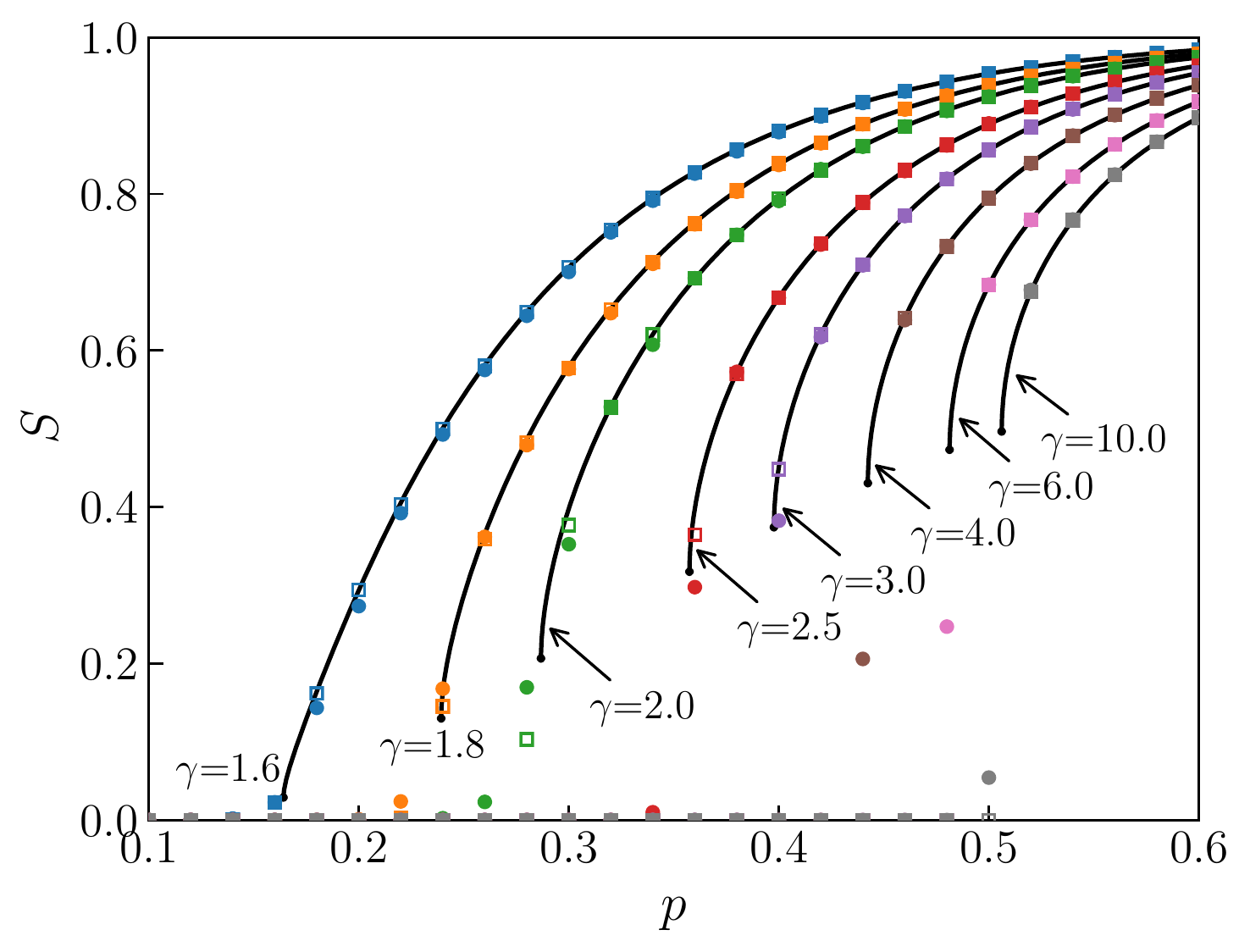}
\caption{
Relative size $S$ of the giant component for three layer multiplex network ($M=3$) with identically powerlaw distributed layers ($P(q) \sim Aq^{-\gamma}, q \geq 4$), as a function of the fraction $p$ of undamaged edges under random edge removal, for various values of $\gamma$. 
(Left) For $\gamma \leq 1 + 1/(M-1) = 1.5$ the giant component appears from $p=0$ with a continuous transition. 
Symbols show measurements averaged over 100 networks of $N=10^4$ nodes (circles) and $N=10^7$ nodes (squares).
(Right) For $\gamma$ larger than this limit, the giant component appears with a discontinuous hybrid transition at a finite critical point. Symbols show measurements in networks of $N=10^5$ nodes (circles, 100 realisations) and $N=10^7$ nodes (squares, one realisation). Black solid curves are theoretical calculations\protect\footnotemark.
\label{M3_small}
\label{M3_small2}}
\end{figure}
\footnotetext{Figures reproduced from \citep{baxter2020weak} with permission of the authors.}

Unlike for the continuous transition in two layers, the critical behaviour of the discontinuous hybrid transition, present in three or more layers, doesn't change when the degree distributions are heavy tailed, except in the most extreme case. It always consists of a discontinuity with square root singularity above the transition point. The degree distribution can, however, have a significant effect on the size of the discontinuity and the critical point. When the degree distribution decays more slowly than $q^{-[1+1/(M-1)]}$ (in the symmetric case), the hybrid transition finally disappears, becoming a continuous transition with zero threshold.

For symmetric uncorrelated layers, i.e. each layer having the same degree distribution, Equation (\ref{powerlaw}), 
recalling the expansion of $G(1-x)$, Equation (\ref{Gexpansion}),
when $\gamma > 2$, for $Z\ll1$ we have
\begin{equation}\label{psi_gam2plus}
Z \cong \left\{  \langle q\rangle Z  - A\Gamma(1-\gamma) Z^{\gamma-1} \right\}^{M-1}\,. 
\end{equation}

Applying the condition for the location of the hybrid transition Equation (\ref{hybrid_crit}), and assuming that $Z \neq 0$, we find
\begin{equation}\label{Zc_gam2plus}
Z_c = \left\{
\frac{\langle q\rangle_c(M-2)} {A\Gamma(1-\gamma)  \left[(M-1)(\gamma-1)-1\right]}\right\}^{1/(\gamma-2)}
\end{equation}
hence
\begin{equation}\label{Zc_gam2plus}
S_c = \left\{
\frac{\langle q\rangle_c(M-2)} {A\Gamma(1-\gamma)  \left[(M-1)(\gamma-1)-1\right]}\right\}^{M/[(M-1)(\gamma-2)]}\,,
\end{equation}
where the critical point $\langle q\rangle_c$ can be found by substituting Equation (\ref{Zc_gam2plus}) into Equation (\ref{psi_gam2plus}):
\begin{multline}
\langle q\rangle_c = 
\left[ \frac{A\Gamma(1-\gamma)[(M-1)(\gamma-1)-1]}{M-2}\right]^{(M-2)/[(M-1)(\gamma-1)-1]}
\\ \times
\left[\frac{[(M-1)(\gamma-1)-1]}{(M-1)(\gamma-2)} \right]^{(M-1)(\gamma-2)/[(M-1)(\gamma-1)-1]}
\end{multline}
which is finite for all $\gamma > 2$.

We plot the size of the giant component $S$ as a function of mean degree in symmetric three layer multiplex networks, for several values of $\gamma$, in Figure \ref{M3_large}. Although the height of the discontinuity reduces as $\gamma$ approaches $2$, it does not disappear. Simulation results are also shown, reproduced from \citep{baxter2020weak}. The agreement is again excellent. Note the increased finite size effects for smaller $\gamma$.

When $1 < \gamma < 2$, again it is not guaranteed that the self consistency equations will hold. However, by comparing with extensive simulations, it was found in \citep{baxter2020weak}, reproduced here in Figure \ref{M3_small2}, 
that the results are indeed accurate.
One finds that the height of the hybrid transition decreases with decreasing $\gamma$, and tends to zero at $\gamma = 1+1/(M-1)$, while the critical point remains finite, as we show in Figure \ref{M3_small2}. 
Below this limit, the threshold is zero and the transition is continuous.

We again apply random damage, using the undamaged fraction of  edges $p$ as the control parameter.
Remembering that the damage does not change 
the asymptotic behaviour of the degree distribution, but reduces the amplitude according to Equation (\ref{Ap}),
the self-consistency equation for $Z$ becomes:
\begin{equation}\label{psib}
Z = \left\{ -A_p\Gamma(1-\gamma) Z^{\gamma-1} -B Z \right\}^{M-1}\,. 
\end{equation}
Note that $\Gamma(1-\gamma)$ is negative in this situation.

Applying the criterion for the hybrid transition point, Equation (\ref{hybrid_crit}), the value of $Z$ above the transition is:
\begin{equation}\label{Zc_general}
Z_c = \left\{
\frac{-A_p\Gamma(1-\gamma)  \left[(M-1)(\gamma-1)-1\right]}{B(M-2)} \right\}^{1/(2-\gamma)}\,.
\end{equation}
This tends to zero at $\gamma = 1 + 1/(M-1)$, and hence the height of the discontinuity, $S_c$, becomes zero at this point.
Writing $\gamma = 1 + 1/(M-1) + \delta$, we can rewrite $Z_c$ near $\delta =0$ as
\begin{align}
Z_c &\cong \left[
\frac{-A_p(M-1)}{B(M-2)} \Gamma\left( \frac{-1}{M-1}\right)
\delta
\right]^{(M-1)/(M-2)}\label{Zc_Mlayers}
\\
&= \left[
\frac{-A_1(M-1)}{B(M-2)} \Gamma\left(\! \frac{-1}{M-1}\!\right)
\delta
\right]^{(M-1)/(M-2)}
p_c^{1/(M-2)} \label{Zc_Mlayersa}
\,.
\end{align}

Substituting Equation  (\ref{Zc_Mlayersa}) back into Equation  (\ref{psib}) gives 
\begin{align}
p_c  \cong \left[-A_1\Gamma\!\left(\frac{-1}{\!M\!-\!1}\!\right)\right]^{-(M-1)}
\left[
    \frac{(M-1)}{B(M-2)} 
\delta
\right]^{-\delta(M\!-\!1)^2/(M\!-\!2)} .
\label{pc_general}
\end{align}

In the limit $\delta \to 0$ the critical point tends to the constant value
\begin{align}
p_c^* =\!\left[-A_1\Gamma\!\left(\frac{-1}{\!M\!-\!1}\!\right)\right]^{-(M-1)}\label{pc_limit}
\end{align}
which depends on the degree distribution only through the amplitude $A$.
That is, the size of the discontinuity $S_c$ tends to zero at $\gamma = 1+1/(M-1)$, yet the critical point $p_c$ remains finite, see Figure \ref{M3_small2}.
For example, for $M=3$, for the distribution used in Figs. \ref{M3_small2} and \ref{M3_small},
\begin{equation}
p_c \to \frac{1}{[A\Gamma(-0.5)]^2} \approx 0.0905... 
 \, \, \text{ as } \gamma \to 1.5^+\,.
\end{equation}
This point is marked with a black circle in Fig. \ref{M3_small}.

Finally, using  that $S = Z^{M/(M-1)}$ in the symmetric case, we have, for edge removal:
\begin{align}\label{Sc_Mlayers_bond}
S_c  
&\cong
\left[
\frac{  (M-1) }{ B(M-2)}\delta
\right]^{M/(M-2)}.
\end{align}

Alternatively, if we apply damage by randomly removing sites, retaining them with probability $p$, 
the self-consistency equation for small $Z$ becomes
\begin{equation}\label{psib_bond}
Z \cong p\left\{ -A\Gamma(1-\gamma) Z^{\gamma-1} -B Z \right\}^{M-1} \equiv \Psi(Z)\,.
\end{equation}
Applying again the condition for the hybrid transition point Eq. (\ref{hybrid_crit}) gives us the same result for $Z_c$, Equation (\ref{Zc_Mlayers}).
Substituting back into Equation (\ref{psib}) also gives the same expression for the critical point, Equation (\ref{pc_general}.
Finally, 
using that, for site removal, $S = p(Z/p)^{M/(M-1)}$ gives:
\begin{align}\label{Sc_Mlayers}
S_c  \cong  \left[
-A\Gamma\left(\frac{-1}{M-1}\right)
\right]^{2(M-1)/(M-2)}
\left[
\frac{  (M-1) }{ B(M-2) }\delta
\right]^{M/(M-2)}\,.
\end{align}

\medskip

We may find the scaling of $S$ with respect to $p-p_c^*$ as we approach this limit from above (that is, approaching the point where the transition becomes continuous)
Let us substitute $\gamma \to [1+ 1/(M-1) ]^+$ into  Equation (\ref{psib}), and rearrange to write $p-p_c^*$ in terms of $Z$. Note that to do so we need to take into account Equation (\ref{pc_limit}).
We find that $Z \sim (p-p_c^*)^{(M-1)/(M-2)}$. Then we may use that $S = Z^{M(M-1)}$ to obtain the limiting scaling relation
\begin{align}
S \cong \left[\frac{B}{(M-1)p_c^*} \right](p-p_c^*)^{M/(M-2)}\,.
\end{align}
Thus, the exponent $\beta(M=3)=3$, while $\beta(M\to \infty) \to 1$.
Note that making the same derivation for the case of site removal leads to an identical exponent.

Below $\gamma = 1+1/(M-1)$ the transition is continuous, with critical point $p_c=0$.
Keeping only the leading order in Equation  (\ref{psib}) and solving for $Z$ we obtain
\begin{align}
Z &= [-A_1\Gamma(1-\gamma)]^{(M-1)/[1-(M-1)(\gamma-1)]} 
 p^{(\gamma-1)(M-1)/[1-(M-1)(\gamma-1)]}
,
\label{610}
\end{align}
giving
\begin{align}
S &= Z^{M/(M-1)}
[-A_1\Gamma(1{-}\gamma)]^{M/[1-(M-1)(\gamma-1)]}p^{(\gamma-1)M/[1-(M-1)(\gamma-1)]}
.
\label{630}
\end{align}

One may alternatively apply random damage to the nodes, retaining a fraction $p$ of them (site percolation), from Equation  (\ref{psib}) we have
\begin{equation}
Z = p[-A_1\Gamma(1-\gamma)]^{M-1}Z^{(M-1)(\gamma-1)},
\label{600b}
\end{equation}
so that
\begin{align}
Z &= [-A_1\Gamma(1-\gamma)]^{(M-1)/[1-(M-1)(\gamma-1)]} p ^{1/[1-(M-1)(\gamma-1)]}
\label{610b}
\end{align}
and hence
\begin{align}
S =& pZ^{M/(M-1)}
\nonumber \\
=& [-A_1\Gamma(1-\gamma)]^{M/[1-(M-1)(\gamma-1)]}
p ^{[2M-1-(M-1)^2(\gamma-1)]/\{(M-1)[1-(M-1)(\gamma-1)]\}}.
\label{630b}
\end{align}

In either case, in this range of $\gamma$ the exponent of $p$ [in Equation (\ref{610b}) or (\ref{630b})] is positive, so $S$ grows as a power of $p$ from $p_c = 0$. 

The growth exponent is smaller for smaller $\gamma$, that is, $S$ grows more rapidly. As $\gamma$ increases, the growth slows, becoming arbitrarily slow in the limit $\gamma \to [1+ 1/(M-1) ]^-$. That is, the transition is of infinite order in this limit.

This is illustrated for $M=3$ layers in Figure \ref{M3_small}, for which the limiting value is $\gamma = 1.5$.
 Finite size effects are strongest close to $\gamma=1+1/(M-1)$, but we see that nevertheless, measurements of finite networks approach the analytical values for $S$ as the size of the network increases, thus reinforcing the validity of the use of self consistency equations in this range of $\gamma$.

%
The interval in which the hybrid transition is absent,  $1<\gamma < 1 + 1/(M-1)$, becomes increasingly small as the number of layers increases. This region vanishes as $M\to\infty$. 

\subsubsection{Non-symmetric layers}

Again, one may obtain results for the asymmetric case, with different $\gamma$ exponents in each layer, by similar methods \citep{baxter2020weak}. Note that in three or more layers, the transition is generally of the discontinuous hybrid type, whose critical behaviour is always the same, not depending on the different values of $\gamma$ exponents in each layer.
The most interesting consideration is then, for which ranges of the $\gamma$s is this transition absent (giving instead a continuous transition)?

For illustrative purposes, let us consider the case $M=3$.
Let us first consider the case that all powerlaw exponents are less than two, $1 < \gamma_a,\gamma_b,\gamma_c < 2$.
When $Z_a, Z_b$ and $Z_c$ are small, we may consider only the leading term in the generating function expansions (ignoring all coefficients):
\begin{eqnarray}
Z_a &\sim& Z_b^{\gamma_b-1} Z_c^{\gamma_c-1}
\,  
\nonumber
\\[3pt]
Z_b &\sim& Z_a^{\gamma_a-1} Z_c^{\gamma_c-1}
\,,  
\nonumber
\\[3pt]
Z_c &\sim& Z_a^{\gamma_a-1} Z_b^{\gamma_b-1}\,.
\label{572}
\end{eqnarray}
From this system of equations we get 
\begin{equation}
Z_a \sim Z_a^{(\gamma_a-1)(\gamma_b-1)+\gamma_a\gamma_b(\gamma_c-1)/\gamma_c}
.
\label{574}
\end{equation}
and similar expressions for $Z_b$ and $Z_c$, which may be found by cycling the subscripts.

The  transition becomes continuous if the exponent of the right-hand side of Equation ~(\ref{574}) is smaller than one. This leads to the following condition for the absence of the discontinuity: 
\begin{eqnarray}
&&
2(\gamma_a{-}1)(\gamma_b{-}1)(\gamma_c{-}1)
\nonumber
\\[3pt]
&&
\!\!\!\!\!\!\!\!\!\!\!\!\!\!\!\!\!\! + (\gamma_a{-}1)(\gamma_b{-}1) + 
(\gamma_b{-}1)(\gamma_c{-}1) + (\gamma_c{-}1)(\gamma_a{-}1) < 1 
.
\label{576}
\end{eqnarray}

If one of the layers, layer $a$, say, has $2 < \gamma_a < 3$, while the other two layers have $1< \gamma_b, \gamma_c < 2$, the expansion for $Z_a$ now has a leading linear term.
We then have:
\begin{align}
Z_a &\sim Z_b^{\gamma_b-1} Z_c^{\gamma_c-1}
\,  
\nonumber
\\
Z_b &\sim Z_a Z_c^{\gamma_c-1}
\,,  
\nonumber
\\
Z_c &\sim Z_a Z_b^{\gamma_b-1}\,.
\label{572alt}
\end{align}
This leads to the condition
\begin{eqnarray}
&&
3(\gamma_b{-}1)(\gamma_c{-}1) + (\gamma_b{-}1) + (\gamma_c{-}1) < 1 
.\label{condition_onegt2}
\end{eqnarray}
For a given $\gamma_b$ between $1$ and $2$, Equation (\ref{condition_onegt2}) gives us a condition for $\gamma_c$:
\begin{align}
\gamma_c < 2\gamma_b/(3\gamma_b-2)
.
\end{align}
the right-hand side of which is always between $1$ (when $\gamma_b=2$) and $2$ (when $\gamma_b=1$).
In other words, the for condition for the disappearance of the discontinuity can also be 
satisfied with one $\gamma >2$.

On the other hand, if, say $\gamma_b >2$ as well, the condition on $\gamma_c$ is:
\begin{eqnarray}
&&
4(\gamma_c{-}1) + 1 < 1 
.
\end{eqnarray}
which cannot be satisfied.




\newpage

\section{Conclusions}\label{conclusions}

Multi-layer networks have lately received significant attention, as a powerful representation of complex systems consisting of multiple interconnected subsystems.
When the functioning of entities in a system requires support, supply or connection in multiple layers, or, equivalently when sites in different layers are dependent on one another in order to function, the system becomes more fragile (vulnerable to failures). Understanding this fragility is therefore an important practical consideration.
The study of multi-layer networks (or multiplex networks) is also interesting theoretically, as they demonstrate a variety of exotic critical phenomena.

The generalisation of percolation to multiplex networks is not unique. While percolation based on mutually connected clusters has received significant study, our aim here is to draw attention to the alternative definition of weak multiplex percolation. 
Which rule one should use depends on the details of the system under study. The two rules may have significantly different behaviour, especially in two layer systems. In this Element we have summarised progress to date in understanding this interesting percolation process.

In two layers there is a continuous transition with quadratic growth of the giant component above the critical point.
In three or more layers we find a discontinuous hybrid transition, which appears in network problems with activation thresholds or constraints (e.g. $k$-core, bootstrap percolation, mutually connected clusters). This is because these constraints allow for the growth of “subcritical” clusters which activate in avalanches under perturbations, and which diverge in size at the critical point, provoking the discontinuity in the size of the giant component. 

Highly heterogeneous network structures, i.e. so called scale-free network layers in which lower moments of the degree distribution diverge, have a strong effect on these critical phenomena. Characterising the degree distribution by the asymptotic powerlaw decay exponent $\gamma$, we show that in two symmetric layers the growth exponent of the giant component above the critical point (beta exponent) becomes non-integral for $\gamma < 3$, and the threshold is zero below $\gamma = 2$.
In three layers, in contrast to other network percolation processes, the character of the transition is not affected until we reach extreme values of $\gamma$. The threshold and the height of the discontinuity of the hybrid transition are both finite above $\gamma = 1+1/(M-1)$. Results in this extreme region of degree distributions, for which even the mean degree diverges with system size, are only able to be verified through simulations in large systems, with careful attention to the convergence as the system size increases. This was done in \citep{baxter2020weak}.
Below  $\gamma = 1+1/(M-1)$ the discontinuity disappears and the critical point becomes zero. The region where this occurs diminishes rapidly with increasing number of layers.

Weak multiplex percolation provides a rich and surprising range of critical behaviour. This simple rule for percolation, for which the status of a node can be decided purely from the status of its neighbours, may be appropriate for a variety of complex systems involving local supply of multiple resources, communication or other activation processes for which the important feature is local connectivity. 

\appendix  

\section{Effect of edge removal on degree distributions}
\label{sa1}

Let us consider a degree distribution with a powerlaw tail, that is
\begin{align}
P(q) \cong A_1q^{-\gamma}
\end{align}
for large $q$.

Under random damage to edges (edge removal), each edge is retained with probability $p$. 
Some nodes will lose all of their edges, and more generally the number of nodes with small degrees will increase, while the number of nodes with large degrees will decrease, however the powerlaw exponent is unchanged. In other words the degree distribution after damage is asymptotically $A_pq^{-\gamma}$ with $A_p < A_1$.

For large $q$, the fraction of vertices with degrees $q'>qp$ in the damaged network should be equal to the the fraction of vertices with degrees $q'>q$ in the original network. (This can be used to show that the damaged network must have the same powerlaw exponent as the undamaged network.)
Calculating these two probabilities by integrating the degree distributions, we find
\begin{equation}
A_p q^{1-\gamma} p^{1-\gamma} \cong A_1 q^{1-\gamma}
.
\label{a20}
\end{equation}
So 
\begin{equation}
A_p  \cong A_1 p^{\gamma-1} 
.
\label{a30}
\end{equation}

%


\section{Generating functions}
\label{sa2}

The generating function of the degree distribution $P(q)$ is defined as 
\begin{equation}
G(x) = \sum_q P(q) x^q
. 
\label{a40}
\end{equation}
In the calculations we give here, the self consistency equations may be written in terms of $G(1-Z)$.
For a Poisson degree distribution with mean degree $\langle q \rangle \equiv c$, 
\begin{align}
G(1-Z) &= e^{-cZ} , \\
G'(1-Z) &= c e^{-cZ} 
. 
\label{a50}
\end{align}

For a scale-free degree distribution, we here for simplicity use a pure power law $P(q) = Aq^{-\gamma}$, beginning from a minimum degree $q_0$. Thus
\begin{align}
G(1-Z) = \sum_{q\geq q_0} Aq^{-\gamma}(1-Z)^q
\end{align}
We may approximate the summation by an integral
\begin{align}
G(1-Z) \approx \int_{q_0}^{\infty}Aq^{-\gamma} e^{-qZ} dq\,
 = A Z^{\gamma-1}\int_{Zq_0}^{\infty}y^{-\gamma} e^{-y} dy\,.
\end{align}
where we have written $y = qZ$.

Integrating by parts twice gives 
\begin{align}\label{Gexp1}
G(1-x) &\cong 1 + B x + A \Gamma(1-\gamma) x^{\gamma-1}+ \mathcal{O}(x^2).
\end{align}
where the coefficient $B$ of the linear term is given by
\begin{equation}\label{B}
B = -\left\{ \sum_q q[P(q) - Aq^{-\gamma}] - \zeta(\gamma-1) \right\}\,.
\end{equation}
For $\gamma > 2$, $B = -\langle q\rangle$. For $1 < \gamma < 2$, $B$ remains finite but depends on the specific form of the degree distribution.

The term in order $\gamma-1$ is either the leading, the second or the third term depending on the value of $\gamma$. Keeping only the leading two terms in $Z$ (after the constant), we thus have the following expansions for $G(1-Z)$ for small $Z$:
\begin{itemize}
\item If $\gamma>3$,
\begin{equation}
G(1-Z) \cong 1 - \langle q \rangle Z + \frac{1}{2} \langle q(q-1) \rangle Z^2.
\label{a60}
\end{equation}
\item If $2<\gamma<3$, then
\begin{equation}
G(1-Z) \cong 1 - \langle q \rangle Z + A \Gamma(1-\gamma) Z^{\gamma-1}
.
\label{a70}
\end{equation}
Note that $\Gamma(1-\gamma)$ is positive in this range.
\item If $1<\gamma<2$, 
\begin{equation}
G(1-Z) \cong 1 + A \Gamma(1-\gamma) Z^{\gamma-1} + BZ.
\label{a80}
\end{equation}
\end{itemize}
Note that here $\Gamma(1-\gamma)$ is now negative.



\bibliography{general_interdependent_networks}

\section*{Acknowledgements}

This work was developed within the scope of the project i3N, UIDB/50025/2020 and UIDP/50025/2020, financed by national funds through the FCT/MEC. This work was also supported by National Funds through FCT, I. P. Project No. IF/00726/2015. R. A. d. C. acknowledges the FCT Grants No. SFRH/BPD/123077/2016 and No. CEECIND/04697/2017.


\end{document}